\documentclass[preprintnumbers, prd, twocolumn, showpacs, floatfix, preprintnumbers, 
letterpaper, 
superscriptaddress,nofootinbib]{revtex4-2}
\usepackage{amsmath}
\usepackage{amsfonts}
\usepackage{amssymb}
\usepackage{mathtools}
\usepackage{bm}
\usepackage{xcolor}

\usepackage{graphicx}

\usepackage{color}
\usepackage{relsize}

\newcommand{\be}{\begin{equation}}
\newcommand{\ee}{\end{equation}}
\newcommand{\bea}{\begin{eqnarray}}
\newcommand{\eea}{\end{eqnarray}}

\begin{document}

\title{Dynamics of dark energy in a scalar-vector-torsion theory }

\author{Manuel Gonzalez-Espinoza}
\email{manuel.gonzalez@upla.cl}
\affiliation{Departamento de Matem\'atica, F\'isica y Computaci\'on, Facultad de Ciencias Naturales y Exactas, Universidad de Playa Ancha, Subida Leopoldo Carvallo 270, Valpara\'iso, Chile.}

\author{Giovanni Otalora}
\email{giovanni.otalora@academicos.uta.cl}
\affiliation{Departamento de F\'isica, Facultad de Ciencias, Universidad de Tarapac\'a, Casilla 7-D, Arica, Chile}

\author{Yoelsy Leyva}
\email{yoelsy.leyva@academicos.uta.cl }
\affiliation{Departamento de F\'isica, Facultad de Ciencias, Universidad de Tarapac\'a, Casilla 7-D, Arica, Chile}

\author{Joel Saavedra}
\email{joel.saavedra@pucv.cl}
\affiliation{Instituto de F\'{\i}sica, Pontificia Universidad Cat\'olica de 
Valpara\'{\i}so, 
Casilla 4950, Valpara\'{\i}so, Chile}

\date{\today}

\begin{abstract} 
We study the cosmological dynamics of dark energy in a scalar-vector-torsion theory. The vector field is described by the cosmic triad and the scalar field is of the quintessence type with non-minimal coupling to gravity. The coupling to gravity is introduced through the interaction between the scalar field and torsion, where torsion is defined in the context of teleparallel gravity. We derive the full set of field equations for the Friedmann–Lema\^{i}tre–Robertson–Walker space-time background and obtain the associated autonomous system.  We obtain the critical points and their stability conditions, along with the cosmological properties of them. Thus, we show that the thermal history of the universe is successfully reproduced. Furthermore, new scaling solutions in which the scalar and vector field densities scale in the same way as the radiation and matter background fluids have been found. Finally, we also show that there exist new attractor fixed points whose nature is mainly vectorial, and which can explain the current accelerated expansion and therefore the dark energy-domination.  

\end{abstract}


\maketitle

\section{Introduction}\label{Introduction}

Currently, one of the biggest puzzles in cosmology is called dark energy \cite{Riess:1998cb,Perlmutter:1998np}. It is believed that dark energy is responsible for the late-time acceleration of the Universe and constitutes $68 \%$ of the total energy density, along with the other $32\%$ associated to dark matter \cite{Zwicky:1933gu}. The simplest  explanation is related to the $\Lambda$CDM model, but it is plagued with theoretical problems, such as the severe fine tuning problem related to its energy scale, the so-called cosmological constant problem \cite{Weinberg:1988cp,Carroll:2000fy,Padilla:2015aaa}. Furthermore, recently some tensions with increasing statistical significant have been found when comparing the $\Lambda$CDM model predictions for cosmological observables by using the cosmic microwave background (CMB) data and the values obtained from independent local measurements \cite{Abdalla:2022yfr,DiValentino:2020zio,DiValentino:2020vvd,Heisenberg:2022lob}. For instance, the value for the Hubble constant today $H_{0}$ estimated from Planck data using the $\Lambda$CDM model is $4.0 \ \sigma$ to  $6.3 \ \sigma$ below current local measurements \cite{Riess:2019qba}, and also the clustering amplitude $\sigma_8$ value deduced from Planck data by using the $\Lambda$CDM model is above that obtained from low-redshift observations \cite{Heymans:2020gsg,Nunes:2021ipq,Heisenberg:2022lob,Heisenberg:2022gqk}. 

As a consequence of these latest findings, the interest in modified gravity models to explain dark energy has increased \cite{DiValentino:2021izs, Clifton:2011jh,Nojiri:2006ri}.  On the one hand, the torsional modified gravity theories are constructed as extensions of the so-called Teleparallel Equivalent of General Relativity, or just Teleparallel Gravity (TG) \cite{Einstein,TranslationEinstein,Early-papers1,Early-papers2,Early-papers3,Early-papers4,Early-papers5,Early-papers6,JGPereira2,AndradeGuillenPereira-00,Arcos:2005ec,Pereira:2019woq,Aldrovandi-Pereira-book}, whose best-known example is $f(T)$ gravity \cite{Bengochea:2008gz,Linder:2010py,Li:2011wu}. For instance, from the perspective of effective field theories (EFT), in Ref. \cite{Yan:2019gbw} the authors have shown that some models of $f(T)$ gravity can efficiently fit observations solving both $H_{0}$ and $\sigma_{8}$ tensions simultaneously. Also, the generalized scalar-torsion $f(T,\phi)$ gravity theories have been studied \cite{Hohmann:2018rwf,Gonzalez-Espinoza:2020azh}.
These include a wide family of theories such as $f(T)$ gravity plus scalar field \cite{Yerzhanov:2010vu,Chakrabarti:2017moe,Rezazadeh:2015dza,Goodarzi:2018feh,Bamba:2016gbu}, nonminimally coupled scalar-torsion theories \cite{Geng:2011aj,Geng:2011ka,Xu:2012jf,Wei:2011yr,Otalora:2013tba,Otalora:2013dsa,Otalora:2014aoa,Skugoreva:2014ena,Jarv:2015odu,Gonzalez-Espinoza:2019ajd}, and their extensions for a nonlinear coupling to torsion \cite{Gonzalez-Espinoza:2020azh,Gonzalez-Espinoza:2020jss,Gonzalez-Espinoza:2021qnv,Gonzalez-Espinoza:2021mwr}. 
Notice that scalar fields are very common in high energy physics and cosmology \cite{Faraoni:2004pi}. Additionally, a nonminimal coupling to gravity 
can arise naturally either due to quantum corrections to the scalar potential \cite{Linde:1982zj}, or due to renormalizability requirements \cite{Freedman:1974gs,Freedman:1974ze,Birrell:1982ix}. Furthermore, these $f(T,\phi)$ gravity theories have shown a rich structure by explaining the dynamics of dark energy \cite{Skugoreva:2014ena,Gonzalez-Espinoza:2020jss, Gonzalez-Espinoza:2021mwr,Leon:2022oyy,Duchaniya:2022fmc,Duchaniya:2022hiy} and inflation \cite{Gonzalez-Espinoza:2020azh,Gonzalez-Espinoza:2021qnv,Leyva:2021fuo} (see also \cite{Cai:2015emx} and references therein). 

In addition to scalar field models, we also have vector field models to explain the current accelerated expansion of the universe. For instance, in the context of generalized Proca theories, a massive vector field breaking the $U(1)$ gauge symmetry is introduced and its time-dependent component has been shown to provide us with an asymptotic de Sitter attractor \cite{DeFelice:2016yws,DeFelice:2016uil,Nakamura:2019phn,DeFelice:2020icf}. However, if the field is canonical and minimally coupled, its motion equation becomes trivial and then it cannot be a source of dark energy \cite{Koivisto:2008xf}.  Let us note that these latter vector theories belong to the class of time-like vector models. On the other hand, space-like vectors have also been considered to model inflation and dark energy \cite{Ford:1989me,Burd:1991ew,Armendariz-Picon:2004say,Koivisto:2008xf,Gomez:2020sfz}. Although, space-like models are typically known to generate a highly anisotropic universe this issue can be solved in some special cases. For instance, one way is by assuming a large number of randomly oriented vectors fields that on the average yields an isotropic background cosmology \cite{Golovnev:2008cf}.  Another way is by considering a set of three identical vector fields for each spatial direction, the so-called cosmic triad which is also consistent with the background symmetry \cite{Armendariz-Picon:2004say}. 

In the present paper, we  study the cosmological dynamics of dark energy in a scalar-vector-torsion theory, with the vector representing the cosmic triad and the scalar field is non-minimally coupled to torsion. The plan of this paper is the following: In section \ref{Intro_TG} we briefly introduce the basics concepts of TG. In Section \ref{model} we present the total action of the model and the field equations. In Section \ref{cosmo_dyna} we obtain the cosmological equations and define the effective dark energy. In Section \ref{phase_space} we perform a phase-space analysis for the model, obtaining the autonomous system with the corresponding critical points and their stability conditions. In Section \ref{Num_Res} we corroborate our previous analytical results by numerically solving the field equations. Finally, in Section \ref{conclusion_f} we summarize the results obtained.

\section{Brief introduction to teleparallel gravity}\label{Intro_TG}
Teleparallel gravity (TG) is an equivalent description of General Relativity (GR), but based on torsion and not curvature \cite{Early-papers5,Early-papers6,Aldrovandi-Pereira-book,Pereira:2019woq}. It is a gauge theory for the translation group \cite{Aldrovandi-Pereira-book,JGPereira2,Arcos:2005ec}. The dynamical variable is the tetrad field which is related to the spacetime metric via
\be
g_{\mu \nu}=\eta_{A B} e^{A}_{~\mu} e^{B}_{~\nu}, 
\ee where $\eta _{AB}^{}=\text{diag}\,(-1,1,1,1)$ is the Minkowski tangent space metric. 

The spin connection of TG is defined as
\be
\omega^{A}_{~B \mu}=\Lambda^{A}_{~D}(x) \partial_{\mu}{\Lambda_{B}^{~D}(x)},
\label{spin_TG}
\ee where $\Lambda^{A}_{~D}(x)$ is a local (point-dependent) Lorentz transformation. For this connection one finds that the curvature tensor vanishes identically
\be
R^{A}_{~B \mu\nu}=\partial_{\mu} \omega^{A}_{~B\nu}-\partial_{\nu}{\omega^{A}_{~B \mu}}+\omega^{A}_{~C \mu} \omega^{C}_{~B \nu}-\omega^{A}_{~C \nu} \omega^{C}_{~B \mu}=0.
\ee This is why it is called purely inertial connection or simply flat connection. Moreover, in the presence of gravity this connection gives a non-vanishing torsion tensor
\be
T^{A}_{~~\mu \nu}=\partial_{\mu}e^{A}_{~\nu} -\partial_{\nu}e^{A}_{~\mu}+\omega^{A}_{~B\mu}\,e^{B}_{~\nu}
 -\omega^{A}_{~B\nu}\,e^{B}_{~\mu}.
\ee 
Using the spin connection one can construct a spacetime-indexed linear connection which is written as
\be
\Gamma^{\rho}_{~~\nu \mu}=e_{A}^{~\rho}\partial_{\mu}e^{A}_{~\nu}+e_{A}^{~\rho}\omega^{A}_{~B \mu} e^{B}_{~\nu}.
\ee It is the so-called Weitzenb\"{o}ck connection. By introducing the contortion tensor 
\begin{equation}  \label{Contortion}
 K^{\rho}_{~~\nu\mu}= \frac{1}{2}\left(T^{~\rho}_{\nu~\mu}
 +T^{~\rho}_{\mu~\nu}-T^{\rho}_{~~\nu\mu}\right),
\end{equation} and the purely spacetime form of the torsion tensor
$T^{\rho}_{~~\mu \nu}=e_{A}^{~\rho} T^{A}_{~~\mu \nu}$
one can verify the following relation 
\be
\Gamma^{\rho}_{~~\nu \mu}=\bar{\Gamma}^{\rho}_{~~\nu \mu}+K^{\rho}_{~~\nu \mu},
\label{RelGamma}
\ee where $\bar{\Gamma}^{\rho}_{~~\nu \mu}$ is the Levi-Civita connection and then 
\be
T^{\rho}_{~~\mu \nu}=\Gamma^{\rho}_{~~\nu \mu}-\Gamma^{\rho}_{~~\mu \nu}.
\ee

The action of TG is quadratic in the torsion tensor and it is given by \cite{Aldrovandi-Pereira-book}
\be
S=-\frac{1}{2 \kappa^2} \int{d^{4}x e ~T},
\ee where $e=\det{(e^{A}_{~\mu})}=\sqrt{-g}$, and $T$ is the torsion scalar, which is defined as
\be
T= S_{\rho}^{~~\mu\nu}\,T^{\rho}_{~~\mu\nu},
\label{ScalarT}
 \ee
with
\begin{equation} \label{Superpotential}
 S_{\rho}^{~~\mu\nu}=\frac{1}{2}\left(K^{\mu\nu}_{~~~\rho}+\delta^{\mu}_{~\rho} \,T^{\theta\nu}_{~~~\theta}-\delta^{\nu}_{~\rho}\,T^{\theta\mu}_{~~~\theta}\right)\,,
\end{equation} the so-called super-potential. 

Interestingly enough, using Eq. \eqref{RelGamma} one can verify that the torsion scalar $T$ of the Weitzenb\"{o}ck connection is related to the curvature scalar $R$ of Levi-Civita connection through
\be
T=-R+2 e^{-1} \partial_{\mu}(e T^{\nu \mu}_{~~~\nu}),
\label{Equiv} 
\ee and therefore, TG and GR are equivalent at the level of field equations. Nevertheless, in order to modify gravity we can start either from GR or TG and the resulting theories do not necessarily have to be equivalent. Thus, in the context of TG, one can modify gravity either by adding a non-minimally coupled matter field such as a scalar field \cite{Geng:2011aj,Otalora:2013tba,Otalora:2013dsa,Otalora:2014aoa,Skugoreva:2014ena}, or considering into the action non-linear terms in the torsion scalar as occurs in $f(T)$ gravity \cite{Bengochea:2008gz,Linder:2010py,Li:2011wu,Gonzalez-Espinoza:2018gyl}. In this way, these torsion-based modified gravity theories belong to new classes of theories that do not have any curvature-based equivalent. In addition, a rich phenomenology has been found which has given rise 
to a fair number of articles in cosmology of early and late-time Universe \cite{Cai:2015emx}.

\section{Scalar-Vector-Torsion theory}\label{model}
We start from the action
\bea
&& S=-\int d^{4}{x} e\Bigg[\frac{T}{2\kappa^2}+F(\phi) T+\frac{1}{2} \partial_{\mu}{\phi}\partial^{\mu}{\phi}+V_1(\phi)+\nonumber\\
&& \sum_{a=1}^{3}\left(\frac{1}{4}\mathcal{F}^{a}_{~\mu \nu}\mathcal{F}^{a \mu \nu}+V_2(A^{a2})\right)\Bigg]+S_{m}+S_{r},
\eea 

where $\mathcal{F}^{a}_{~\mu \nu}=\partial_{\mu} A^{a}_{\nu}-\partial_{\nu}A^{a}_{\mu}$ and $A^{a2}=g^{\mu \nu}A^{a}_{\mu} A^{a}_{\nu}$. $S_{m}$ is the action of matter. $S_{r}$ is the action of radiation.

Varying the action with respect to the tetrad field we obtain the field equations
\bea
&& \left(\frac{1}{2 \kappa^2}+F(\phi)\right) G_{\mu \nu}+S_{\mu \nu}^{\ \ \rho} \partial_{\rho}{F}+\nonumber\\
&& \frac{1}{4}g_{\mu \nu}\left[\frac{1}{2}\partial^{\rho}{\phi}\partial_{\rho}{\phi}+V_1(\phi)+\sum_{a=1}^{3}\left(\frac{1}{4} \mathcal{F}^{a}_{~\sigma \rho} \mathcal{F}^{a \sigma \rho}+V_2(A^{a2})\right)\right]-\nonumber\\
&& \frac{1}{4}\partial_{\mu}{\phi}\partial_{\nu}{\phi}-\frac{1}{4}\sum_{a=1}^{3}\left[\mathcal{F}^{a}_{~\rho \mu} \mathcal{F}^{a \rho}_{~~\nu}+2 A^{a}_{\mu}A^{a}_{\nu} \frac{dV_2}{dA^{a2}}\right]\nonumber\\
&& =\frac{1}{4} T^{(m)}_{~\mu \nu}+\frac{1}{4} T^{(r)}_{~\mu \nu},
\eea where $G^{\mu}_{~\nu}=e_{A}^{~\mu} G^{A}_{~\nu}$ is the Einstein tensor with \cite{Aldrovandi-Pereira-book}
\bea
&& G_{A}^{~\mu}\equiv e^{-1}\partial_{\nu}\left(e e_{A}^{~\sigma} S_{\sigma}^{~\mu\nu}\right)-e_{A}^{~\sigma} T^{\lambda}_{~\rho \sigma}S_{\lambda}^{~\rho \mu}+\nonumber\\
&& e_{B}^{~\lambda} S_{\lambda}^{~\rho \mu}\omega^{B}_{~A \rho}+\frac{1}{4}e_{A}^{~\mu} T.
\eea  

Varying with respect to the scalar field $\phi$ we find the motion equation
\be
\nabla_{\mu}{\partial^{\mu}{\phi}}-F_{,\phi} T-V_{1,\phi}=0, 
\ee where $\nabla_{\mu}$ is the covariant derivative associated to the Levi-Civita connection. 
Finally, varying with respect to the cosmic triad $A^{a}_{\mu}$
\bea
\nabla_{\nu}\mathcal{F}^{a \nu \mu}-2 \frac{dV_2}{dA^{a2}}A^{a \mu}=0.
\eea

Below we study dynamics of the fields in a cosmological background. 

\section{Cosmological dynamics}\label{cosmo_dyna}
In order to study cosmology in the above scalar-vector-torsion theory we choose the background tetrad field
\be
\label{veirbFRW}
e^A_{~\mu}={\rm
diag}(1,a,a,a),
\ee
which corresponds to a flat Friedmann-Robertson-Walker
(FRW) universe with metric 
\begin{equation}
ds^2=-dt^2+a^2\,\delta_{ij} dx^i dx^j \,,
\label{FRWMetric}
\end{equation}
where $a$ is the scale factor which is a function of the cosmic time $t$. 

We also assume the ansatz

\be
A^{a}_{\mu}=\delta^{a}_{\mu} A(t) a(t).
\ee In this way, the three vectors have the same time-dependent length, $A^{a2}=A(t)^2$, being that they point in three mutually orthogonal spatial directions \cite{Armendariz-Picon:2004say}.

Thus, the modified Friedmann equations are given by
\bea
6 H^2 \left( \frac{1}{2 \kappa ^2} + F \right)&=& \frac{3}{2} \left(A H+\dot{A}\right)^2+V_1+3 V_2+\frac{\dot{\phi }^2}{2} \nonumber\\
&+&\rho _m+\rho _r,\label{00}\\
-4 \dot{H} \left(\frac{1}{2 \kappa ^2} +  F\right)&=& 2 \left(A H+\dot{A}\right)^2 + \dot{\phi }^2 + 4 H \dot{\phi } F_{,\phi} \nonumber\\
&+&2 A^2 V_{2,A^{2}}+\rho _m+\frac{4 \rho _r}{3},\label{ii}
\eea along with the motion equation for the fields $\phi$ and $A$
\bea
\ddot{\phi }+6 H^2 F_{,\phi}+3 H \dot{\phi }+V_{1,\phi}&=&0,\\
\ddot{A}+2 A H^2+3 \dot{A} H+A \dot{H}+2 A V_{2,A^{2}} &=&0,
\eea respectively.

Thus, as in Ref. \cite{Copeland:2006wr}, the Friedmann equations \eqref{00} and \eqref{ii}  can be rewritten in their standard form 
\bea
\label{SH00}
&& \frac{3}{\kappa^2} H^2=\rho_{de}+\rho_{m}+\rho_{r},\\
&& -\frac{2}{\kappa^2} \dot{H}=\rho_{de}+p_{de}+\rho_{m}+\frac{4}{3}\rho_{r},
\label{SHii}
\eea where the effective energy and pressure densities are defined as  
\bea
\rho_{de}&= \rho_{v}+\rho_{s} =& \frac{3}{2} \left(A H+\dot{A}\right)^2+3 V_2\nonumber \\ 
&& +\frac{\dot{\phi }^2}{2}+V_1- 6 H^2 F,\\
 p_{de}&= p_{v}+p_{s} =& \frac{1}{2} \left(A H+\dot{A}\right)^2 -3 V_2  +2 A^2 V_{2,A^{2}} \nonumber\\
 && +\frac{\dot{\phi }^2}{2}-V_1 + 6 H^2 F+4 H \dot{\phi } F_{,\phi}+ 4 F \dot{H}. \nonumber\\
&&\eea 
Then, the effective dark energy equation-of-state (EOS) parameter is
\begin{equation}
w_{de}=\frac{p_{de}}{\rho _{de}}.
\label{wDE1}
\end{equation}
For these definitions of $\rho_{de}$ and $p_{de}$ one can verify that they satisfy
\begin{eqnarray}
\dot{\rho}_{de}+3H(\rho_{de}+p_{de})=0.
\end{eqnarray} This equation is consistent with the energy conservation law and the fluid evolution equations
\bea
\label{rho_m}
&& \dot{\rho}_{m}+3 H\rho_{m}=0,\\
&& \dot{\rho}_{r}+4 H\rho_{r}=0.
\label{rho_r}
\eea

Also, it is useful to introduce the
total EOS parameter as 
\be
w_{tot}=\frac{p_{de}+p_r}{\rho _{de}+\rho _m+\rho _r},
\label{wtot}
\ee which is related to the deceleration parameter $q$ through
\be
q=\frac{1}
{2}\left(1+3w_{tot}\right).
\label{deccelparam}
\ee Then, the acceleration of the Universe occurs for $q<0$, or equivalently for  $w_{tot}<-1/3$.

Finally, another useful set of cosmological parameters which we can introduce are the so-called standard density parameters  
\bea
&& \Omega_{m}\equiv\frac{\kappa^2 \rho_{m}}{3 H^2},\:\:\:\: \Omega_{de}\equiv\frac{\kappa^2 \rho_{de}}{3 H^2}=\Omega_{v}+\Omega_{s},\:\:\:\: \Omega_{r}\equiv \frac{\kappa^2 \rho_{r}}{3 H^2},\nonumber\\
&& \Omega_{v}=\frac{\kappa^2 \rho_{v}}{3 H^2},\:\:\:\: \Omega_{s}=\frac{\kappa^2 \rho_{s}}{3 H^2},
\eea which satisfy the constraint equation  
\be
\Omega_{de}+\Omega_{m}+\Omega_{r}=1.
\ee This latter equation is nothing more than the same energy constraint \eqref{SH00}, but written in terms of the energy density parameters. 

In terms of these energy density parameters the total equation of state can be also written as 
\be
w_{tot}=w_{v}\Omega_{v}+w_{s}\Omega_{s}+\frac{1}{3}\Omega _{r},
\ee  where we have defined 
\be
w_{v}=\frac{p_{v}}{\rho_{v}},\:\:\:\: w_{s}=\frac{p_{s}}{\rho_{s}}, 
\ee which represent the EOS parameters of the vector and the scalar fields, respectively. Let us remember that for non-relativistic matter one has $w_{m}=p_{m}/\rho_{m}=0$, and for relativistic matter $w_{r}=p_{r}/\rho_{r}=1/3$ \cite{amendola2010dark}.

In the next section we perform an exhaustive phase space analysis for this dark energy model. Particularly, we obtain the corresponding autonomous system from the set of cosmological equations \eqref{SH00}, \eqref{SHii}, \eqref{rho_m} and \eqref{rho_r}. 

\section{Phase space Analysis}\label{phase_space}
In order to write the set of the cosmological equations of the model as an autonomous system we introduce the following useful dimensionless variables \cite{Copeland:2006wr}

\begin{table*}[ht]
 \centering
 \caption{Critical points for the autonomous system. }
\begin{center}
\begin{tabular}{c c c c c c c c c}\hline\hline
Name &  $x_{1c}$ & $x_{2c}$ & $x_{3c}$ & $y_{1c}$ & $y_{2c}$ & $u_{c}$ & $\varrho_{c}$  \\\hline
$\ \ \ \ \ \ \ \ a_{R} \ \ \ \ \ \ \ \ $ & $0$ & $0$ & $x_{3 c}$ & $0$ & $0$ & $0$  & $\sqrt{1-3 x_{3 c}^2}$ \\
$\ \ \ \ \ \ \ \ b_{R} \ \ \ \ \ \ \ \ $ & $\frac{2 \sqrt{\frac{2}{3}}}{\lambda _1}$ & $0$ & $x_{3 c}$ & $\frac{2}{\sqrt{3} \lambda _1}$ & $0$ & $0$  & $\frac{\sqrt{\lambda _1^2-3 \lambda _1^2 x_{3 c}^2-4}}{\lambda _1}$ \\
$\ \ \ \ \ \ \ \ c_{M} \ \ \ \ \ \ \ \ $ & $0$ & $0$ & $0$ & $0$ & $0$ & $0$  & $0$ \\
$\ \ \ \ \ \ \ \ d_{M} \ \ \ \ \ \ \ \ $ & $\frac{\sqrt{\frac{3}{2}}}{\lambda _1}$ & $0$ & $0$ & $\frac{\sqrt{\frac{3}{2}}}{\lambda _1}$ & $0$ & $0$  & $0$ \\
$\ \ \ \ \ \ \ \ e_{SM} \ \ \ \ \ \ \ \ $ & $1$ & $0$ & $0$ & $0$ & $0$ & $0$  & $0$ \\
$\ \ \ \ \ \ \ \ f \ \ \ \ \ \ \ \ $ & $0$ & $0$ & $0$ & $\frac{\sqrt{\sigma _1-3 \sigma _1 y_{2 c}^2}}{\sqrt{\lambda _1+\sigma _1}}$ & $y_{2c}$ & $\frac{\lambda _1-3 \lambda _1 y_{2 c}^2}{\lambda _1+\sigma _1}$  & $0$ \\
$\ \ \ \ \ \ \ \ g \ \ \ \ \ \ \ \ $ & $0$ & $0$ & $x_{3c}$ & $\frac{\sqrt{-3 \lambda _2 \sigma _1 x_{3 c}^2+\lambda _2 \sigma _1-\sigma _1}}{\sqrt{\lambda _2 \sigma _1+\lambda _1 \lambda _2}}$ & $\frac{1}{\sqrt{3} \sqrt{\lambda _2}}$ & $\frac{-3 \lambda _1 \lambda _2 x_{3 c}^2-\lambda _1+\lambda _1 \lambda _2}{\lambda _2 \left(\lambda _1+\sigma _1\right)}$  & $0$ \\
$\ \ \ \ \ \ \ \ h \ \ \ \ \ \ \ \ $ & $0$ & $0$ & $0$ & $\frac{\sqrt{\sigma _1}}{\sqrt{\lambda _1+\sigma _1}}$ & $0$ & $\frac{\lambda _1}{\lambda _1+\sigma _1}$  & $0$ \\
$\ \ \ \ \ \ \ \ i \ \ \ \ \ \ \ \ $ & $\frac{\lambda _1}{\sqrt{6}}$ & $0$ & $0$ & $\frac{\sqrt{6-\lambda _1^2}}{\sqrt{6}}$ & $0$ & $0$  & $0$ \\
$\ \ \ \ \ \ \ \ j \ \ \ \ \ \ \ \ $ & $0$ & $0$ & $\frac{\sqrt{\lambda _2-1}}{\sqrt{3} \sqrt{\lambda _2}}$ & $0$ & $\frac{1}{\sqrt{3} \sqrt{\lambda _2}}$ & $0$  & $0$ \\

\\ \hline\hline
\end{tabular}
\end{center}
\label{table1}
\end{table*}

\begin{eqnarray}
x_{1} =& \dfrac{\kappa  \dot{\phi}}{\sqrt{6} H}, \ \ \ \ \ \ x_{2} =& \dfrac{\kappa  \dot{A}}{\sqrt{6} H}, \ \ \ \ \ \ \ x_3 =\frac{\kappa  A}{\sqrt{6}},\nonumber\\ 
y_1 =& \dfrac{\kappa  \sqrt{V_1}}{\sqrt{3} H}, \ \ \ \ \ \
y_2 =& \dfrac{\kappa  \sqrt{V_2}}{\sqrt{3} H}, \ \ \ \ \ \ \  u = -2 \kappa ^2 F,
\nonumber\\
\lambda_1 =& - \dfrac{V_{1,\phi}}{\kappa V_1}, \ \ \ \ \ \ \ \lambda_2 =& - \dfrac{V_{2,A^{2}}}{\kappa^2 V_2}, \ \ \ \ \ \ \ \sigma = - \dfrac{F_{,\phi}}{\kappa F}, \nonumber\\
 \varrho =& \dfrac{\kappa  \sqrt{\rho_r}}{\sqrt{3} H },\ \ \ \ \ \ \ \ \ \ \ \ & \nonumber\\
&& 
\label{var}
\end{eqnarray}
and the constraint equation
\begin{equation}
    \Omega_m + \varrho^2 + u+3 x_{2}^2+3 x_{3}^2+6 x_{2} x_{3}+3 y_{2}^2+x_1^2+y_1^2 = 1,
\end{equation}
Therefore, we obtain the dynamical system
\begin{eqnarray}
\dfrac{d x_1}{d N} &=& -\frac{f_1(x_1,x_2,x_3,y_1,y_2,u,\varrho)}{2 \left(u-1\right)},
\nonumber\\
\dfrac{d x_2}{d N} &=& -\frac{f_2(x_1,x_2,x_3,y_1,y_2,u,\varrho)}{2 \left(u-1\right)}, \nonumber\\
\dfrac{d x_3}{d N} &=& x_2, \nonumber\\
\dfrac{d y_1}{d N} &=& -\frac{y_1 f_3(x_1,x_2,x_3,y_1,y_2,u,\varrho)}{2 \left(u-1\right)}, \nonumber\\
\dfrac{d y_2}{d N} &=& -\frac{y_2 f_4(x_1,x_2,x_3,y_1,y_2,u,\varrho)}{2 \left(u-1\right)}, \nonumber\\
\dfrac{d u}{d N} &=& -\sqrt{6} \sigma _1 u x_1, \nonumber\\
\dfrac{d \varrho}{d N} &=& -\frac{\varrho f_5(x_1,x_2,x_3,y_1,y_2,u,\varrho)}{2 \left(u-1\right)}, \nonumber\\
\dfrac{d \lambda_1}{d N} &=& -\sqrt{6} \left(\Gamma _1-1\right) \lambda _1^2 x_1, \nonumber\\
\dfrac{d \lambda_2}{d N} &=& -12 \left(\Gamma _2-1\right) \lambda _2^2 x_2 x_3, \nonumber\\
\dfrac{d \sigma_1}{d N} &=& -\sqrt{6} \left(\Theta _1-1\right) \sigma _1^2 x_1, \nonumber\\
&&   \label{ODE10}
\end{eqnarray}

where
\begin{eqnarray}
f_1 &=& 2 \sqrt{6} \sigma _1 u x_1^2 +x_1 (3 u-36 \lambda _2 x_3^2 y_2^2+3 x_2^2+3 x_3^2+6 x_2 x_3 \nonumber\\
&-&3 y_1^2-9 y_2^2+\varrho ^2-3)+\sqrt{6} (u-1) \left(\sigma _1 u-\lambda _1 y_1^2\right)+3 x_1^3, \nonumber\\
f_2 &=& x_2 (u (2 \sqrt{6} \sigma _1 x_1+3)-36 \lambda _2 x_3^2 y_2^2+3 x_1^2+9 x_3^2-3 y_1^2 \nonumber\\
&-&9 y_2^2+\varrho ^2-3)+x_3 (2 \sqrt{6} \sigma _1 u x_1-12 \lambda _2 u y_2^2+u
\nonumber\\
&-&36 \lambda _2 x_3^2 y_2^2+3 x_1^2+3 x_3^2+12 \lambda _2 y_2^2-3 y_1^2-9 y_2^2+\varrho ^2-1)\nonumber\\
&+&3 x_2^3+9 x_3 x_2^2, \nonumber\\
f_3&=& u (\sqrt{6} x_1 (\lambda _1+2 \sigma _1)-3)-\sqrt{6} \lambda _1 x_1-36 \lambda _2 x_3^2 y_2^2+3 x_1^2 \nonumber\\
&+&3 x_2^2+3 x_3^2+6 x_2 x_3-3 y_1^2-9 y_2^2+\varrho ^2+3 ,\nonumber\\
f_4 &=& u (12 \lambda _2 x_2 x_3+2 \sqrt{6} \sigma _1 x_1-3)-12 \lambda _2 x_2 x_3-36 \lambda _2 x_3^2 y_2^2 \nonumber\\
&+&3 x_1^2+3 x_2^2+3 x_3^2+6 x_2 x_3-3 y_1^2-9 y_2^2+\varrho ^2+3,\nonumber\\
f_5 &=& 2 \sqrt{6} \sigma _1 u x_1+u-36 \lambda _2 x_3^2 y_2^2+3 x_1^2+3 x_2^2+3 x_3^2 \nonumber\\
&+& 6 x_2 x_3-3 y_1^2-9 y_2^2+\varrho ^2-1 .\nonumber\\
&&
\end{eqnarray}

Using the above set of phase space variables we can also write
\bea
\Omega_{de} &=& u+x_1^2+3 x_2^2+3 x_3^2+6 x_2 x_3+y_1^2+3 y_2^2,\\
\Omega_{m} &=& -u-x_1^2-3 x_2^2-3 x_3^2-6 x_2 x_3-y_1^2-3 y_2^2-\varrho ^2+1. \nonumber\\
&&
\eea

Similarly, the equation of state of dark energy $w_{de}=p_{de}/\rho_{de}$ can be rewritten as
\bea
w_{de} &=& -(2 \sqrt{6} \sigma _1 u x_1+u \varrho ^2-36 \lambda _2 x_3^2 y_2^2+3 x_1^2 \nonumber\\
&+&3 x_2^2+3 x_3^2+6 x_2 x_3-3 y_1^2-9 y_2^2) / (3 (u-1) \nonumber\\
&&(u+x_1^2+3 x_2^2+3 x_3^2+6 x_2 x_3+y_1^2+3 y_2^2)), \nonumber\\
&&
\eea
whereas the total equation of state becomes
\bea
w_{tot}&=& \frac{1}{3-3 u} (2 \sqrt{6} \sigma _1 u x_1-36 \lambda _2 x_3^2 y_2^2+3 x_1^2+3 x_2^2+3 x_3^2\nonumber\\
&+&6 x_2 x_3-3 y_1^2-9 y_2^2+\varrho ^2).
\eea

To obtain an autonomous system from the dynamical system \eqref{ODE10}, we need to define the potentials of both the scalar and vector fields, as well as the non-minimal coupling function.
From now we concentrate in the exponential potential $V_{1} (\phi)\sim e^{-\kappa \lambda_{1} \phi}$ and $V_{2}(A^{2})\sim e^{-\kappa^{2} \lambda_{2} A^{2}}$. {This type of potential is known to allow scaling solutions \cite{Copeland:2006wr}}. {Also, the most natural and simple choice for the nonminimal coupling function, which is compatible with the exponential scalar potential, is represented by the function $F(\phi)\sim e^{-\kappa \sigma \phi}$ \cite{Amendola:1999qq}}. Below, we study the critical points and their stability properties.

\subsection{Critical points}

We obtain the critical points from the conditions $dx_{1}/dN = dx_{2}/dN = dx_{3}/dN = dy_1/dN = dy_2/dN=du/dN=d\varrho/dN=0$. Also, the definitions in Eq. \eqref{var} imply that the values of the critical points are real, with $y_{1}>0$, $y_{2}>0$, and $\varrho>0$. The critical points for system \eqref{ODE10} are shown in Table \ref{table1}, and their values of cosmological parameters in Table \ref{table2}.

The critical point $a_{R}$ represents a scaling radiation era with  $\Omega^{(r)}_{de}=3 x_{3 c}^2$. For $x_{3 c}=0$ we recover the radiation-dominated solution with $\Omega_{r}=1$ and $\omega_{de}=\omega_{tot}=1/3$. Otherwise, to be consistent with the earliest constraint coming from the physics of big bang nucleosynthesis (BBN), $\Omega^{(r)}_{de}<0.045$ \cite{Ferreira:1997hj,Bean:2001wt}, we require $x_{3c}<0.122$. Point $b_{R}$ also corresponds to a scaling radiation with $\Omega^{(r)}_{de}=3 x_{3 c}^2+4/\lambda_{1}^2$. In this case, the radiation-dominated era can only be possible in the limit $x_{3 c}\rightarrow 0$ and $\lambda_{1}\rightarrow \pm \infty$.
On the other hand, the critical point $c_{M}$ is a matter-dominated era with $\Omega_{m}=1$ and $\omega_{de}=\omega_{tot}=0$. Critical point $d_{M}$ describes a scaling matter era with $\Omega^{(r)}_{de}=3/\lambda_{1}^2$ and then the  matter-dominated era is recovered for $\lambda_{1}\rightarrow \pm \infty$. 

Critical point $e_{SM}$ is a dark energy-dominated solution with $\Omega_{de}=1$. However, this solution cannot explain the current accelerated expansion because its equation of state is of the stiff matter type with $\omega_{de}=\omega_{tot}=1$.  Critical points $f$, $g$, $h$ and $j$ are also dark energy-dominated solutions with a de Sitter equation of state $\omega_{de}=\omega_{tot}=-1$. Therefore, these solutions provide accelerated expansion for all the values of parameters. Interestingly enough,  points $f$ and $g$ have contributions from both the scalar and vector fields, while the points $h$ and $j$ only receive contribution from the scalar and vector fields, respectively. Finally, point $i$ is a dark energy-dominated solution with an equation of state of the quintessence type with accelerated expansion for $\lambda_{1}<\sqrt{2}$. This point corresponds to the standard accelerated solution for quintessence models \cite{Copeland:2006wr}.

\subsection{Stability of critical points}\label{Stability}

In this section we study the stability of the critical points. For this we assume time-dependent linear perturbations around each critical point, and from the autonomous system \eqref{ODE10} the corresponding linear perturbation matrix $\mathcal{M}$ is obtained \cite{Copeland:2006wr}. The eigenvalues of $\mathcal{M}$ evaluated at each fixed point give the stability conditions that we are looking for. Thus, we follow the standard classification of the stability properties: (i) Stable node: all the eigenvalues are negative;
(ii) Unstable node: all the eigenvalues are positive; (iii) Saddle point: at least one is positive and the others are negative; (iv) Stable spiral: the determinant of $\mathcal{M}$ is negative, and the real part of all the eigenvalues are negative. Points which are stable node or stable spiral are called attractor points, and these fixed points are reached through the cosmological evolution of the Universe, independently of the initial conditions. 


\begin{table}[hb]
 \centering
 \caption{Cosmological parameters for the critical points in Table \ref{table1}. }
\begin{center}
\begin{tabular}{c c c c c c}\hline\hline
Name &   $\Omega_{de}$ & $\Omega_{m}$ & $\Omega_{r}$ & $\omega_{de}$ & $\omega_{tot}$ \\\hline
$a_{R}$ & $3 x_{3 c}^2$ & $0$ & $1-3 x_{3 c}^2$ & $\frac{1}{3} $ & $\frac{1}{3}$ \\
$b_{R}$ &  $3 x_{3 c}^2+\frac{4}{\lambda _1^2}$ & $0$ & $-3 x_{3 c}^2-\frac{4}{\lambda _1^2}+1$ & $\frac{1}{3}$ & $\frac{1}{3}$ \\
$c_M$ &  $0$ & $1$ & $0$ & $0$ & $0$ \\
$d_M$ &  $\frac{3}{\lambda _1^2}$ & $1-\frac{3}{\lambda _1^2}$ & $0$ & $0$ & $0$ \\
$e_{SM}$ &  $1$ & $0$ & $0$ & $1$ & $1$ \\
$f$ &  $1$ & $0$ & $0$ & $-1$ & $-1$ \\
$g$ &  $1$ & $0$ & $0$ & $-1$ & $-1$ \\
$h$ &  $1$ & $0$ & $0$ & $-1$ & $-1$ \\
$i$ &  $1$ & $0$ & $0$ & $\frac{1}{3} \left(\lambda _1^2-3\right)$ & $\frac{1}{3} \left(\lambda _1^2-3\right)$ \\
$j$ &  $1$ & $0$ & $0$ & $-1$ & $-1$ \\
\\ \hline\hline
\end{tabular}
\end{center}
\label{table2}
\end{table}


Now, we analyze the corresponding eigenvalues for each critical point and their stability conditions.

\begin{itemize}
    \item Point $a_R$ has the eigenvalues
    \begin{eqnarray}
        &\mu_1 = 2, \ \ \ \ \mu_2 = 2, \ \ \ \ \mu_3 = -1, \ \ \ \ \mu_4 = -1,\nonumber\\
         &\mu_5 = 1, \ \ \ \ \mu_6 = 0, \ \ \ \ \mu_7 = 0.
    \end{eqnarray} Since this critical point have positive and negative eigenvalues we can conclude that it is always unstable. 
    \item Point $b_R$ has the eigenvalues
    \begin{eqnarray}
        &\mu_1 = 0, \ \ \ \ \mu_2 = -1, \ \ \ \ \mu_3 = 1, \ \ \ \ \mu_4 = 2,\nonumber\\
         &\mu_5 = -\frac{\sqrt{64-15 \lambda _1^2}+\lambda _1}{2 \lambda _1}, \ \ \ \ \mu_6 = \frac{\sqrt{64-15 \lambda _1^2}-\lambda _1}{2 \lambda _1}, \nonumber\\
         &\mu_7 = -\frac{4 \sigma _1}{\lambda _1}.
    \end{eqnarray} For the same reason as for the point $a_{R}$, the point $b_R$ is also always unstable. 
    \item Point $c_M$ has the eigenvalues
    \begin{eqnarray}
        &\mu_1 = -\frac{3}{2}, \ \ \ \ \mu_2 =\frac{3}{2}, \ \ \ \ \mu_3 = \frac{3}{2}, \ \ \ \ \mu_4 = -1,\nonumber\\
         &\mu_5 = -\frac{1}{2}, \ \ \ \ \mu_6 = -\frac{1}{2}, \ \ \ \ \mu_7 = 0.
    \end{eqnarray} Similarly, this point is always unstable.
    \item Point $d_M$ has the eigenvalues
    \begin{eqnarray}
        &\mu_1 = -1, \ \ \ \ \mu_2 = -\frac{1}{2}, \ \ \ \ \mu_3 = -\frac{1}{2}, \ \ \ \ \mu_4 = \frac{3}{2},\nonumber\\
         &\mu_5 = -\frac{3 \left(\sqrt{24-7 \lambda _1^2}+\lambda _1\right)}{4 \lambda _1}, \ \ \ \ \mu_6 = \frac{3 \left(\sqrt{24-7 \lambda _1^2}-\lambda _1\right)}{4 \lambda _1}, \nonumber\\
         &\mu_7 = -\frac{3 \sigma _1}{\lambda _1}.
    \end{eqnarray} This point is always a saddle point for $-2 \sqrt{\frac{6}{7}}<\lambda_{1}<2 \sqrt{\frac{6}{7}}$.
    \item Point $e_{SM}$ has the eigenvalues
    \begin{eqnarray}
        &\mu_1 = 3, \ \ \ \ \mu_2 = 3, \ \ \ \ \mu_3 = -1, \ \ \ \ \mu_4 = 1,\nonumber\\
         &\mu_5 = 1, \ \ \ \ \mu_6 = 3-\sqrt{\frac{3}{2}} \lambda _1, \ \ \ \ \mu_7 = -\sqrt{6} \sigma _1.
    \end{eqnarray} This point is always a saddle point.
    {\item Point $h$ has the eigenvalues
    \begin{eqnarray}
        &\mu_1 = 0, \ \ \ \ \mu_2 = -1, \ \ \ \ \mu_3 = -2, \ \ \ \ \mu_4 = -2,\nonumber\\
         &\mu_5 = -3, \ \ \ \ \mu_6 = \frac{1}{2} \left(-\sqrt{12 \lambda _1 \sigma _1+9}-3\right),\nonumber\\ 
         &\ \ \ \ \mu_7 = \frac{1}{2} \left(\sqrt{12 \lambda _1 \sigma _1+9}-3\right),
    \end{eqnarray}
    where, 
    \begin{eqnarray}
        \left(\sigma _1<0\land 0<\lambda _1\leq -\frac{3}{4 \sigma _1}\right)\lor \nonumber\\
        \left(\sigma _1>0\land -\frac{3}{4 \sigma _1}\leq \lambda _1<0\right),
    \end{eqnarray}
    to be marginally stable.}
    \item Point $i$ has the eigenvalues
    \begin{eqnarray}
        &\mu_1 = -1, \ \ \ \ \mu_2 = \frac{\lambda _1^2}{2}, \ \ \ \ \mu_3 = \frac{1}{2} \left(\lambda _1^2-6\right), \nonumber\\
         & \mu_4 = \frac{1}{2} \left(\lambda _1^2-4\right), \ \ \ \ \mu_5 = \frac{1}{2} \left(\lambda _1^2-4\right), \ \ \ \ \mu_6 = \lambda _1^2-3, \nonumber\\
         &\mu_7 = -\lambda _1 \sigma _1. \label{cp_i}
    \end{eqnarray} This point is always a saddle point. 
    \item Point $j$ has the eigenvalues
    \begin{eqnarray}
        &\mu_1 = 0, \ \ \ \ \mu_2 = 0, \ \ \ \ \mu_3 = -3, \ \ \ \ \mu_4 = -3,\nonumber\\
         &\mu_5 = -2, \ \ \ \ \mu_6 = -\frac{\sqrt{-\lambda _2 \left(7 \lambda _2-16\right)}+3 \lambda _2}{2 \lambda _2},\nonumber\\ 
         &\ \ \ \ \mu_7 = \frac{\sqrt{-\lambda _2 \left(7 \lambda _2-16\right)}-3 \lambda _2}{2 \lambda _2},
    \end{eqnarray}
    where, 
    \begin{equation}
        1< \lambda_2 \leq \frac{16}{7},
    \end{equation}
    to be marginally stable.
\end{itemize}

The eigenvalues associated to critical {points $f$ and $g$ are quite long, and we add them to appendix \ref{eigenfg}.} We found that at linear order, these critical points are also marginally stable. In Section \ref{Num_Res} we verify numerically that they are indeed stable fixed points.   

Below we perform a numerical integration of the autonomous system to corroborate our analytical results.


\begin{figure}[!tbp]
  \centering
  \begin{minipage}[b]{0.45\textwidth}
    \includegraphics[width=\textwidth, height=4cm]{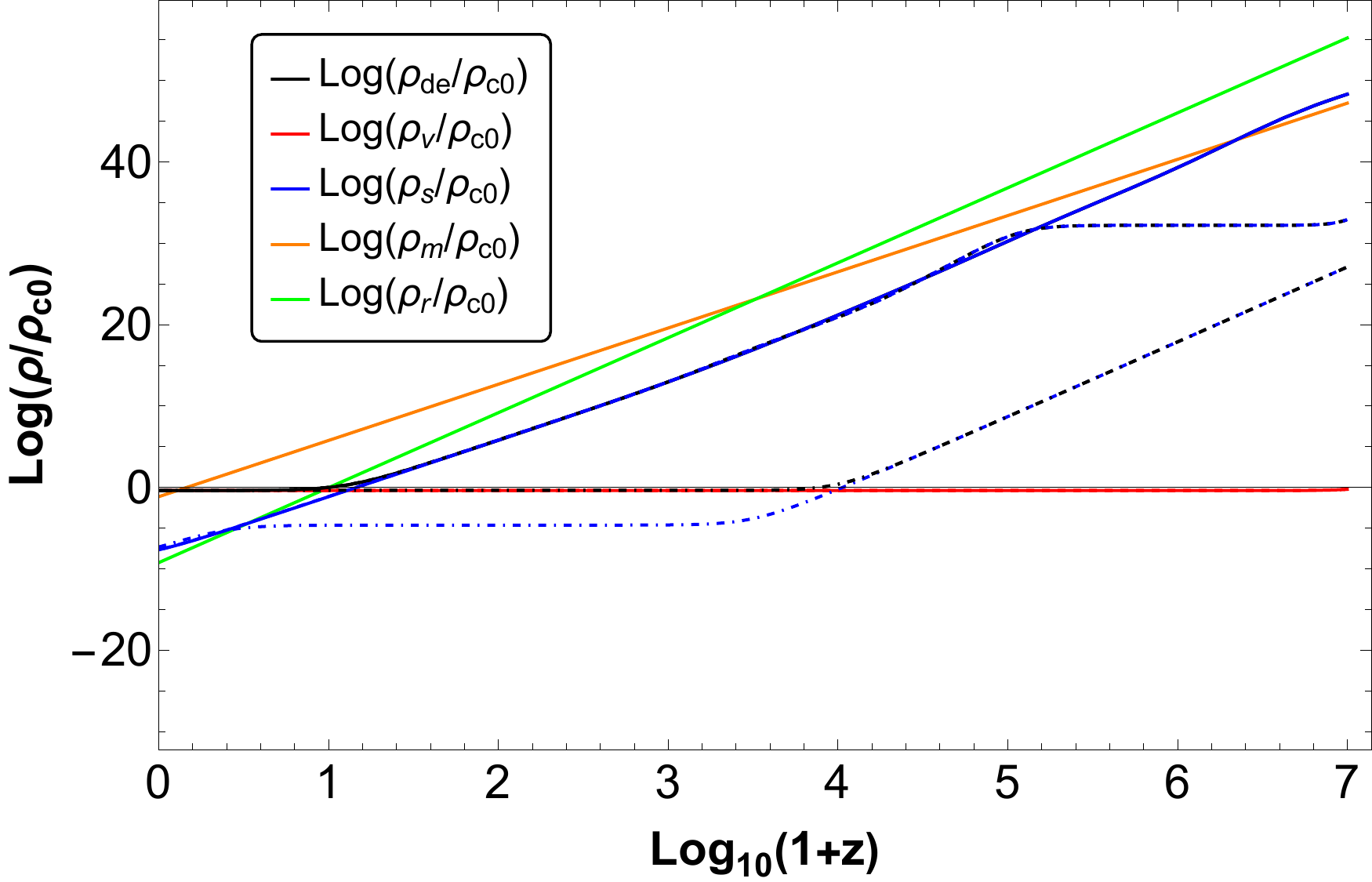}
    \caption{\smaller{We depict the evolution of the energy density of dark energy $\rho_{de}$ (black), scalar field density $\rho_{s}$ (blue), vector field density $\rho_{v}$ (red), dark matter (including baryons) $\rho_m$ (orange) and radiation $\rho_r$ (green) as functions of the redshift z, for $\lambda_1 = 5.5 \times 10^{1}$, $\lambda_2 = 1$ and $\sigma_1 = 10^{-1}$. In particular, solid lines correspond to initial conditions $x_{1 i} = 0.0222681$, $x_{2 i} = 10^{-13}$, $x_{3 i} = 10^{-13}$, $y_{1 i} = 0.0222681$, $y_{2 i} = 0.485 \times 10^{-12}$, $u_{i} = 6 \times 10^{-13}$ and $\varrho_{i} = 0.999339$. Dashed lines correspond to initial conditions $x_{1 i} = 10^{-5}$, $x_{2 i} = 10^{-13}$, $x_{3 i} = 10^{-13}$, $y_{1 i} = 10^{-5}$, $y_{2 i} = 0.485 \times 10^{-12}$, $u_{i} = 6 \times 10^{-13}$ and $\varrho_{i} = 0.99983$. Dot-dashed lines correspond to initial conditions $x_{1 i} = 10^{-13}$, $x_{2 i} = 10^{-13}$, $x_{3 i} = 10^{-13}$, $y_{1 i} = 10^{-13}$, $y_{2 i} = 0.485 \times 10^{-12}$, $u_{i} = 6 \times 10^{-13}$ and $\varrho_{i} = 0.99983$. During the scaling radiation era $a_{R}$ we obtained $\Omega_{de}^{(r)}\approx 9.91736 \times 10^{-4}$, while in the case of $b_{R}$ we found {$\Omega_{de}^{(r)}\approx 1.66852 \times 10^{-3}$}. Also, during the scaling matter era $d_{M}$, we got {$\Omega_{de}^{(m)}\approx 1.00912 \times 10^{-3}$} at redshift $z=50$.\\ \\}} 
    \label{Figura1}
  \end{minipage}
  \hfill
  \begin{minipage}[b]{0.45\textwidth}
    \includegraphics[width=\textwidth, height=4cm]{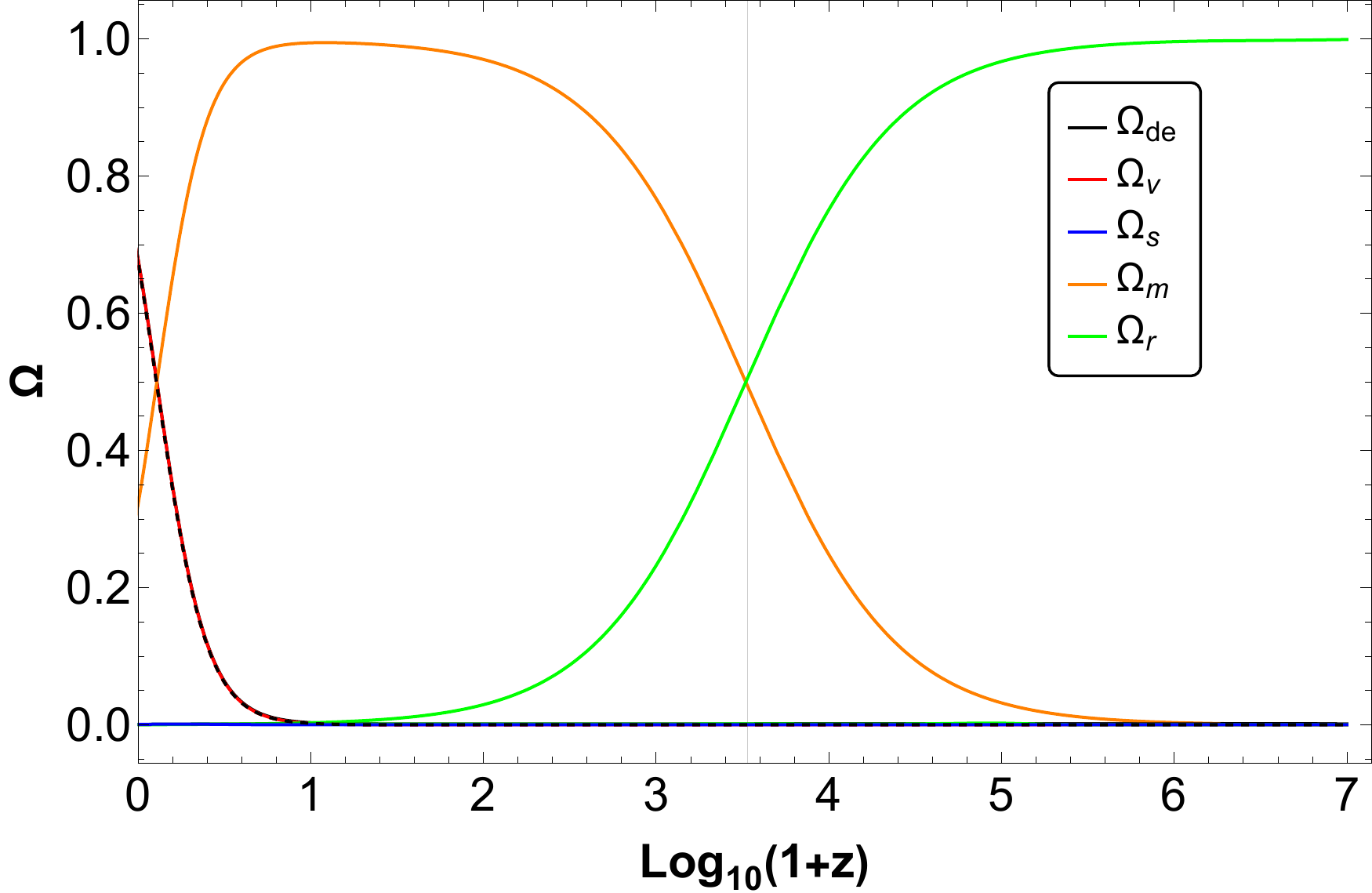}
    \caption{\smaller{The evolution of the fractional energy densities for the same initial conditions used in figure \ref{Figura1}. Where we can observe that the fractional energy density of dark energy (black line) and the fractional energy density related to vector field (red line) are overlapping, and the fractional energy density related to scalar field is around $10^{-4}$ (blue line). Also, the time of radiation-matter equality is around $z \approx 3387$. Finally, at future times, $\Omega_v \approx 0.999946$ and $\Omega_s \approx 5.37922 \times 10^{-5}$.}} 
    \label{Figura2}
  \end{minipage}
\end{figure}

\begin{figure}[!tbp]
  \centering
  \begin{minipage}[b]{0.45\textwidth}
    \includegraphics[width=\textwidth, height=4cm]{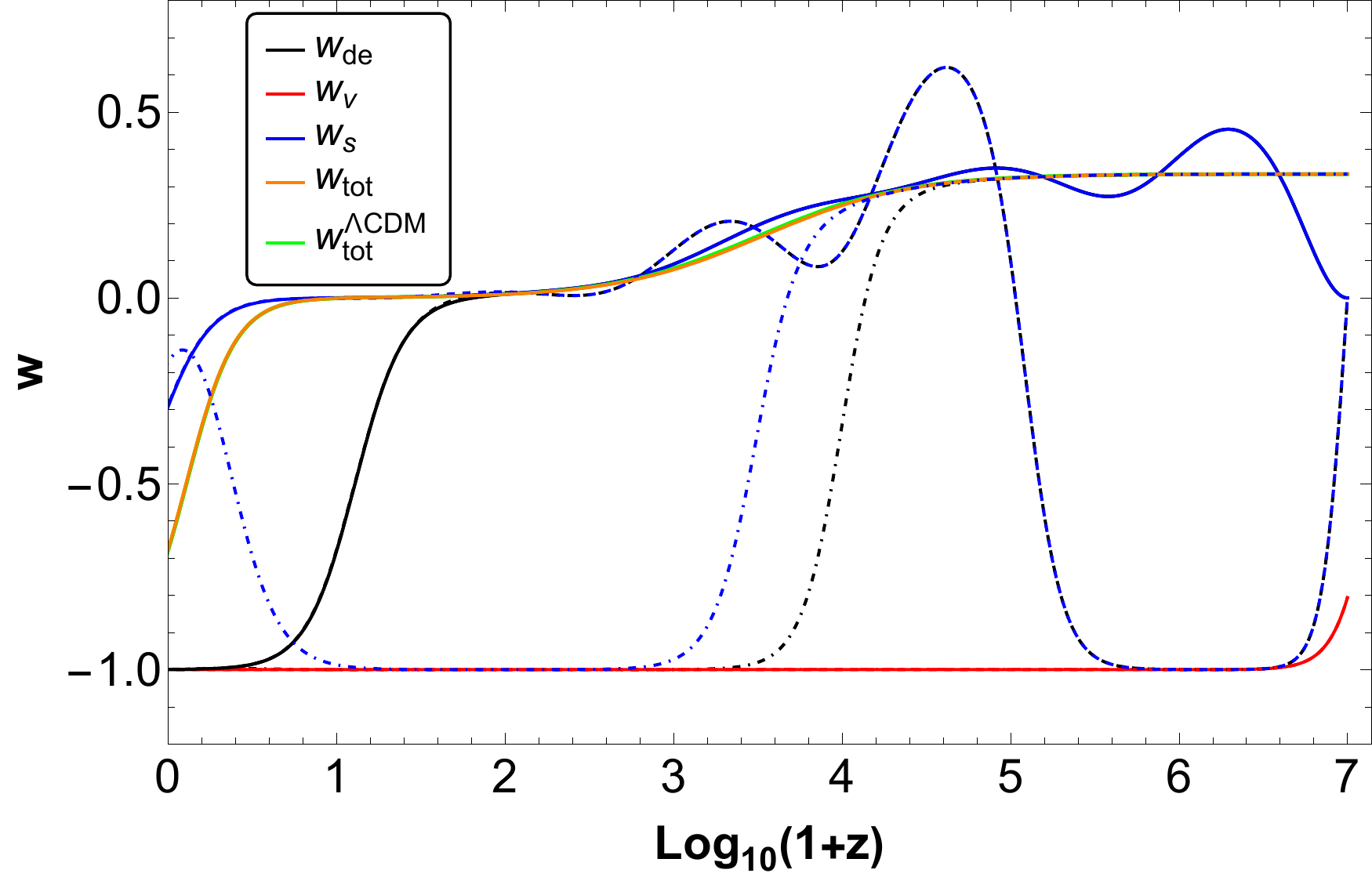}
    \caption{\smaller{We depict the equation of state $w_{tot}$ (orange line), the EOS of dark energy $w_{de}$ (black line), the EOS related to vector field $w_{v}$ (red line), the EOS related to scalar field $w_{s}$ (blue line), and the total EOS of $\Lambda$CDM model (green line) as redshift functions.  Also, we used the same initial conditions of figure \ref{Figura1} to obtain the solid, dashed, and dot-dashed blue lines and red lines. Finally, we obtain the value $w_{de} = -0.999516$ at $z = 0$, in accordance with $w^{(0)}_{de} =-1.028 \pm 0.032$, from Planck data \cite{Akrami:2018odb}.\\ \\}} 
    \label{Figura3}
  \end{minipage}
  \hfill
  \begin{minipage}[b]{0.45\textwidth}
    \includegraphics[width=\textwidth, height=4cm]{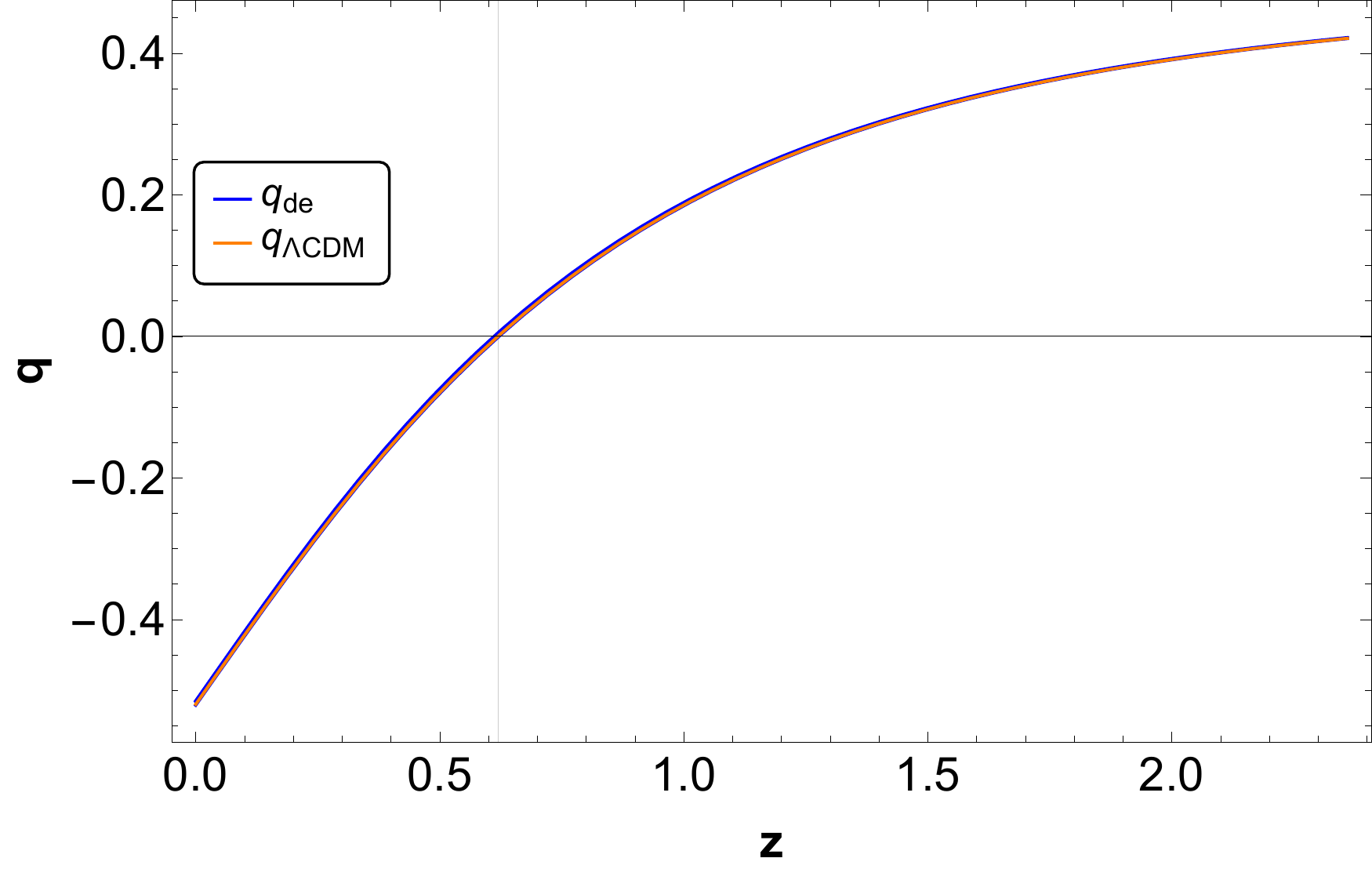}
    \caption{\smaller{We show the evolution of the deceleration parameter $q(z)$, for the same initial conditions used in figure \ref{Figura1}. The transition to dark energy dominance is around $z \approx 0.62$, consistent with current observational data.}} 
    \label{Figura5}
  \end{minipage}
\end{figure}

\begin{figure}[!tbp]
  \centering
  \begin{minipage}[b]{0.45\textwidth}
    \includegraphics[width=\textwidth, height=4cm]{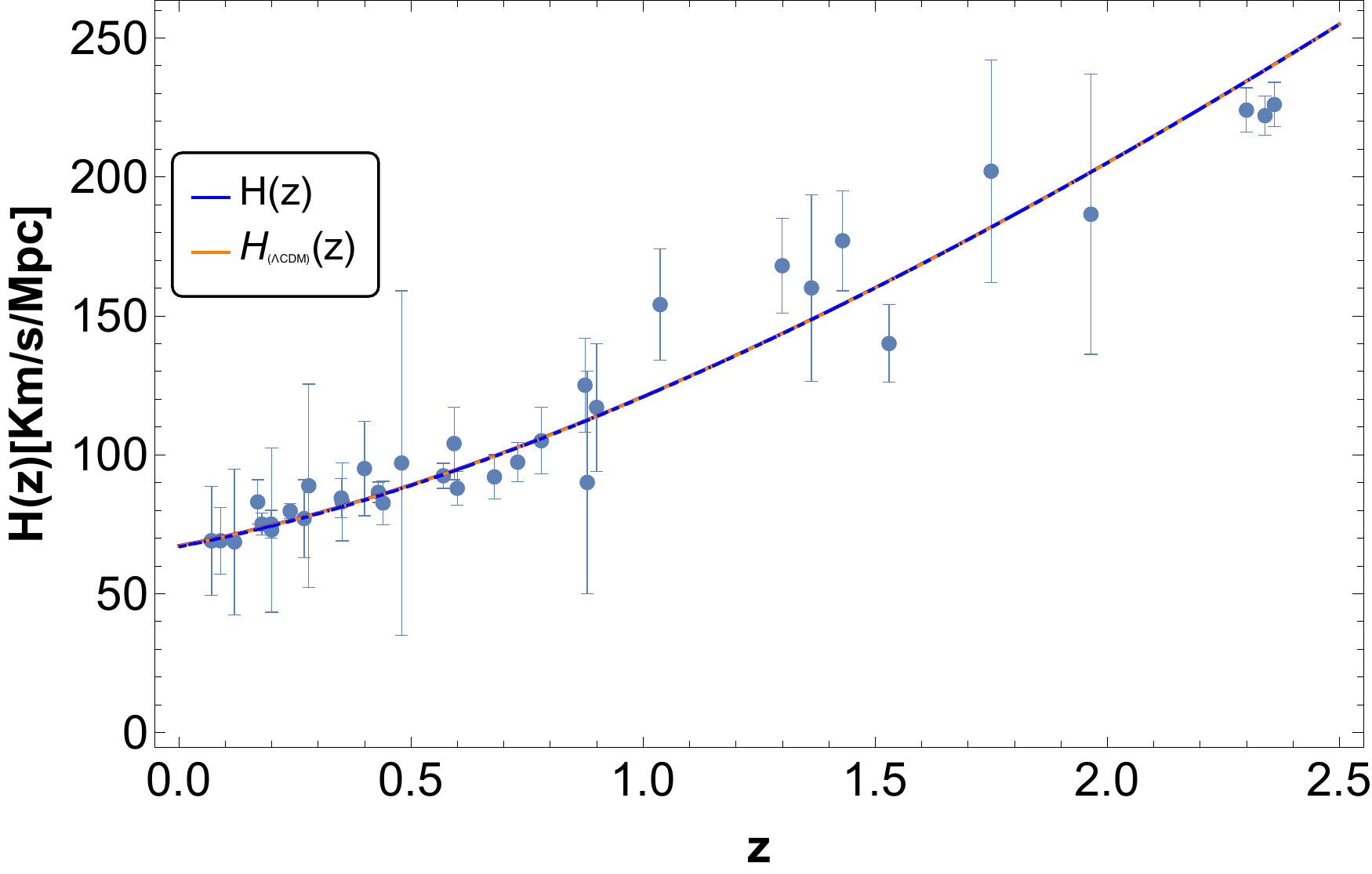}
    \caption{\scriptsize{We present the variation of the Hubble rate, denoted as $H(z)$, in relation to the redshift $z$. The observed data points for the Hubble rate are depicted as blue points, accompanied by their corresponding $1 \sigma$ confidence intervals, as detailed in Table \ref{table:H(z)data}. We utilize identical initial conditions as displayed in FIG \ref{Figura1} to generate the solid, dashed, and dot-dashed blue lines. By comparing the theoretical model predictions represented by these lines with the observed data, we can assess the level of agreement between the empirical measurements of the Hubble rate at different redshifts and the theoretical model.\\}} 
    \label{Figura1H}
  \end{minipage}
  \hfill
  \begin{minipage}[b]{0.45\textwidth}
    \includegraphics[width=\textwidth, height=4cm]{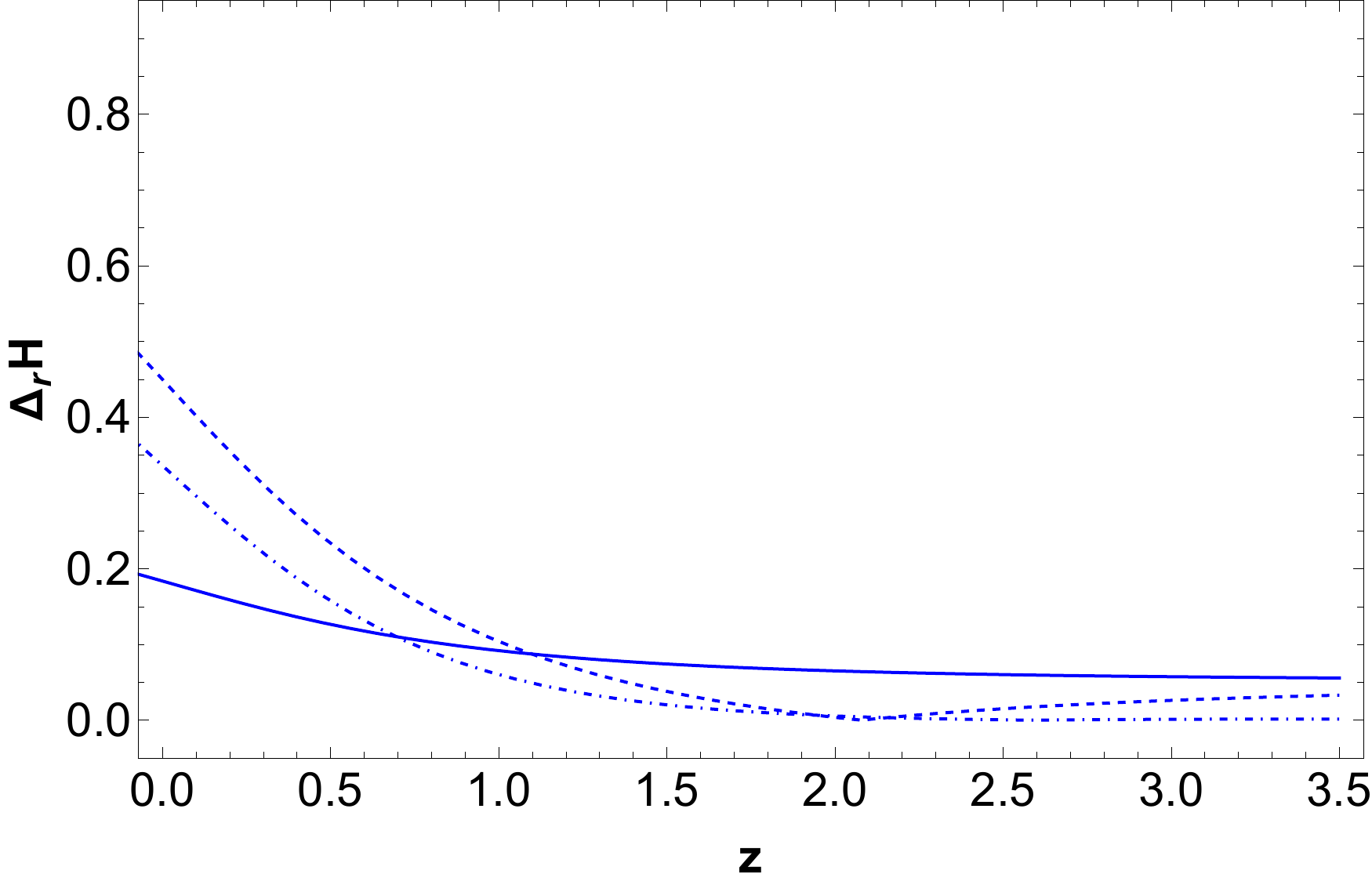}
    \caption{\scriptsize{The figure illustrates the relative difference (expressed as a percentage), denoted as $\Delta_r H(z) = 100 \times \left| H - H_{\Lambda CDM} \right|/H_{\Lambda CDM}$, in comparison to the $\Lambda$CDM model. The solid, dashed, and dot-dashed blue lines are derived from the identical initial conditions presented in FIG. \ref{Figura1}, are utilized in this analysis. By evaluating the relative difference between the Hubble rate $H$ and the Hubble rate of the $\Lambda$CDM model $H_{\Lambda CDM}$, we can examine the deviations from the $\Lambda$CDM model and the associated significance at different redshifts.}} 
    \label{FiguradH1}
  \end{minipage}
\end{figure}


\begin{figure}[!tbp]
	\centering
		\includegraphics[width=0.35\textwidth]{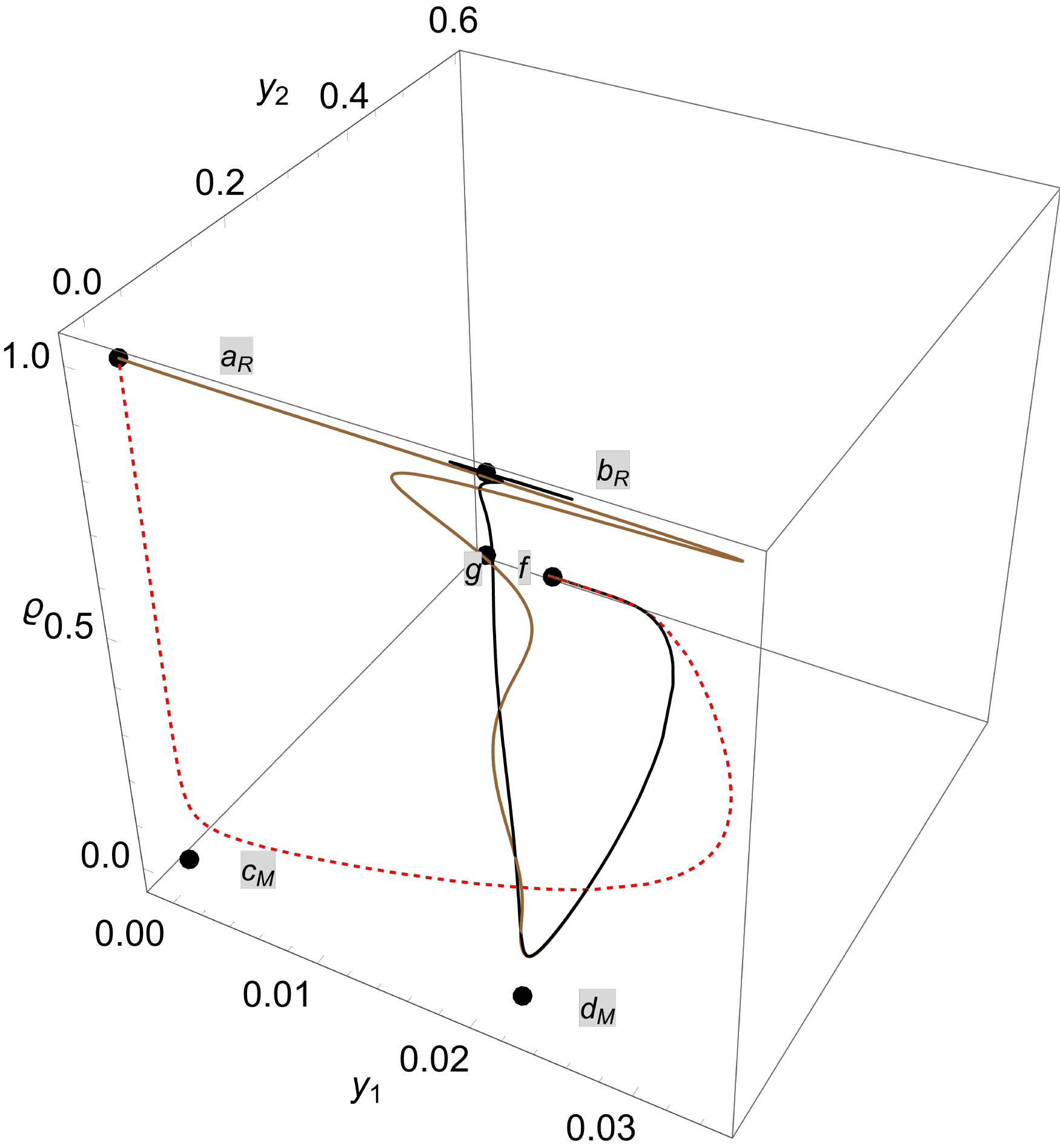}
	\caption{\smaller{ Evolution curves in the phase space for $\lambda_1 = 5.5 \times 10^{1}$, $\lambda_2 = 1$ and $\sigma_1 = 10^{-1}$. Black solid line correspond to initial conditions $x_{1 i} = 0.0222681$, $x_{2 i} = 10^{-13}$, $x_{3 i} = 10^{-13}$, $y_{1 i} = 0.0222681$, $y_{2 i} = 0.485 \times 10^{-12}$, $u_{i} = 6 \times 10^{-13}$ and $\varrho_{i} = 0.999339$. Brown solid line correspond to initial conditions $x_{1 i} = 10^{-5}$, $x_{2 i} = 10^{-13}$, $x_{3 i} = 10^{-13}$, $y_{1 i} = 10^{-5}$, $y_{2 i} = 0.485 \times 10^{-12}$, $u_{i} = 6 \times 10^{-13}$ and $\varrho_{i} = 0.99983$. Red dashed line correspond to initial conditions $x_{1 i} = 10^{-13}$, $x_{2 i} = 10^{-13}$, $x_{3 i} = 10^{-13}$, $y_{1 i} = 10^{-13}$, $y_{2 i} = 0.485 \times 10^{-12}$, $u_{i} = 6 \times 10^{-13}$ and $\varrho_{i} = 0.99983$.}} 
	\label{Figura4}
\end{figure}



\begin{figure}[!tbp]
  \centering
  \begin{minipage}[b]{0.45\textwidth}
    \includegraphics[width=\textwidth, height=4cm]{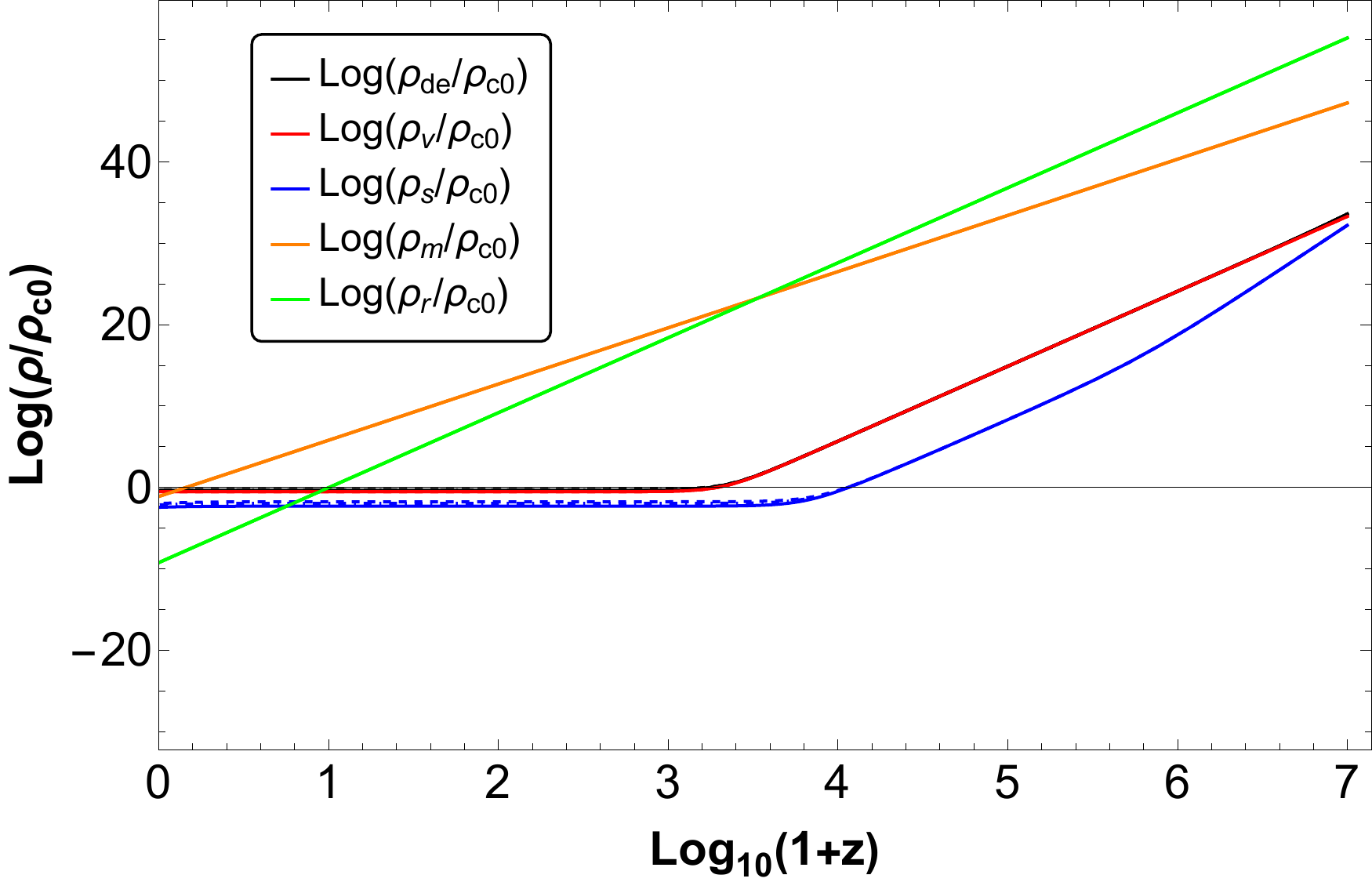}
    \caption{\smaller{We depict the evolution of the energy density of dark energy $\rho_{de}$ (black), scalar field density $\rho_{s}$ (blue), vector field density $\rho_{v}$ (red), dark matter (including baryons) $\rho_m$ (orange) and radiation $\rho_r$ (green) as functions of the redshift z, for $\lambda_1 = 2.2$, $\lambda_2 = 1$ and $\sigma_1 = 1$. In particular, solid lines correspond to initial conditions $x_{1 i} = 10^{-5}$, $x_{2 i} = 10^{-5}$, $x_{3 i} = 10^{-13}$, $y_{1 i} = 3.2 \times 10^{-13}$, $y_{2 i} = 4.5 \times 10^{-13}$, $u_{i} = 4 \times 10^{-13}$ and $\varrho_{i} = 0.99983$. Dashed lines correspond to initial conditions $x_{1 i} = 10^{-5}$, $x_{2 i} = 10^{-5}$, $x_{3 i} = 10^{-13}$, $y_{1 i} = 4.2 \times 10^{-13}$, $y_{2 i} = 4.5 \times 10^{-13}$, $u_{i} = 4 \times 10^{-13}$ and $\varrho_{i} = 0.99983$. Dot-dashed lines correspond to initial conditions $x_{1 i} = 10^{-5}$, $x_{2 i} = 10^{-5}$, $x_{3 i} = 10^{-13}$, $y_{1 i} = 3.8 \times 10^{-13}$, $y_{2 i} = 4.5 \times 10^{-13}$, $u_{i} = 4.0 \times 10^{-13}$ and $\varrho_{i} = 0.99983$. During the scaling radiation era $a_{R}$ we obtained $\Omega_{de}^{(r)}\approx 4.004 \times 10^{-10}$.\\ \\}} 
    \label{Figura6}
  \end{minipage}
  \hfill
  \begin{minipage}[b]{0.45\textwidth}
    \includegraphics[width=\textwidth, height=4cm]{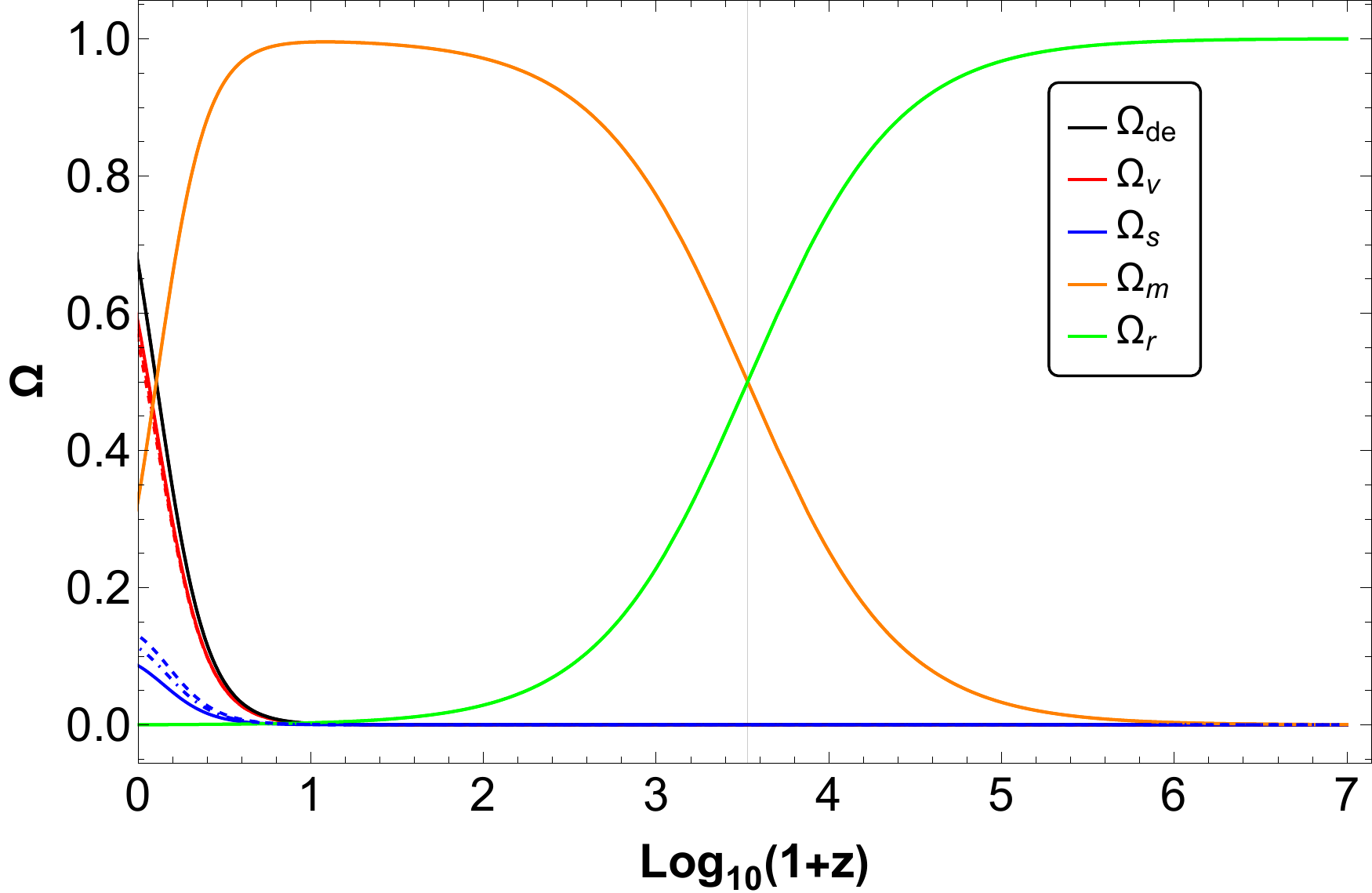}
    \caption{\smaller{The evolution of the fractional energy densities of dark energy (black line), vector field (red line) and  scalar field (blue line), for the same initial conditions used in figure \ref{Figura6}. Also, the time of radiation-matter equality is around $z \approx 3387$.  Finally, at future times, $\Omega_v \approx 0.96571$ and $\Omega_s \approx 3.42898 \times 10^{-2}$.}} 
    \label{Figura7}
  \end{minipage}
\end{figure}

\begin{figure}[!tbp]
  \centering
  \begin{minipage}[b]{0.45\textwidth}
    \includegraphics[width=\textwidth, height=4cm]{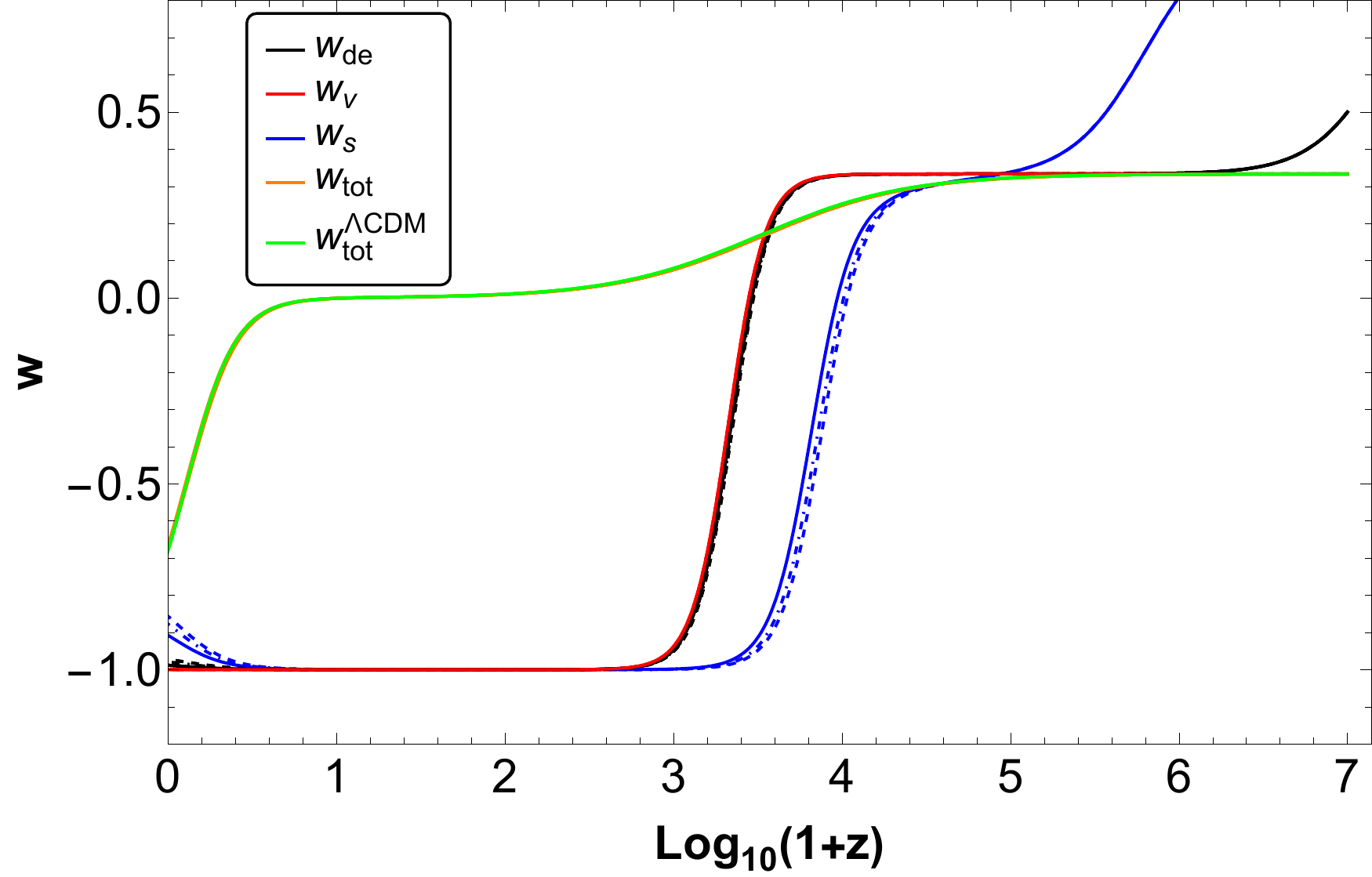}
    \caption{\smaller{We depict the equation of state $w_{tot}$ (orange line), the EOS of dark energy $w_{de}$ (black line), the EOS related to vector field $w_{v}$ (red line), the EOS related to scalar field $w_{s}$ (blue line), and the total EOS of $\Lambda$CDM model (green line) as redshift functions.  Also, we used the same initial conditions of figure \ref{Figura7} to obtain the solid, dashed, and dot-dashed blue lines and red lines. Finally, we obtain the value $w_{de} = -0.996$ at $z = 0$, in accordance with $w^{(0)}_{de} =- 1.028 \pm 0.032$, from Planck data \cite{Akrami:2018odb}.\\ \\}} 
    \label{Figura8}
  \end{minipage}
  \hfill
  \begin{minipage}[b]{0.45\textwidth}
    \includegraphics[width=\textwidth, height=4cm]{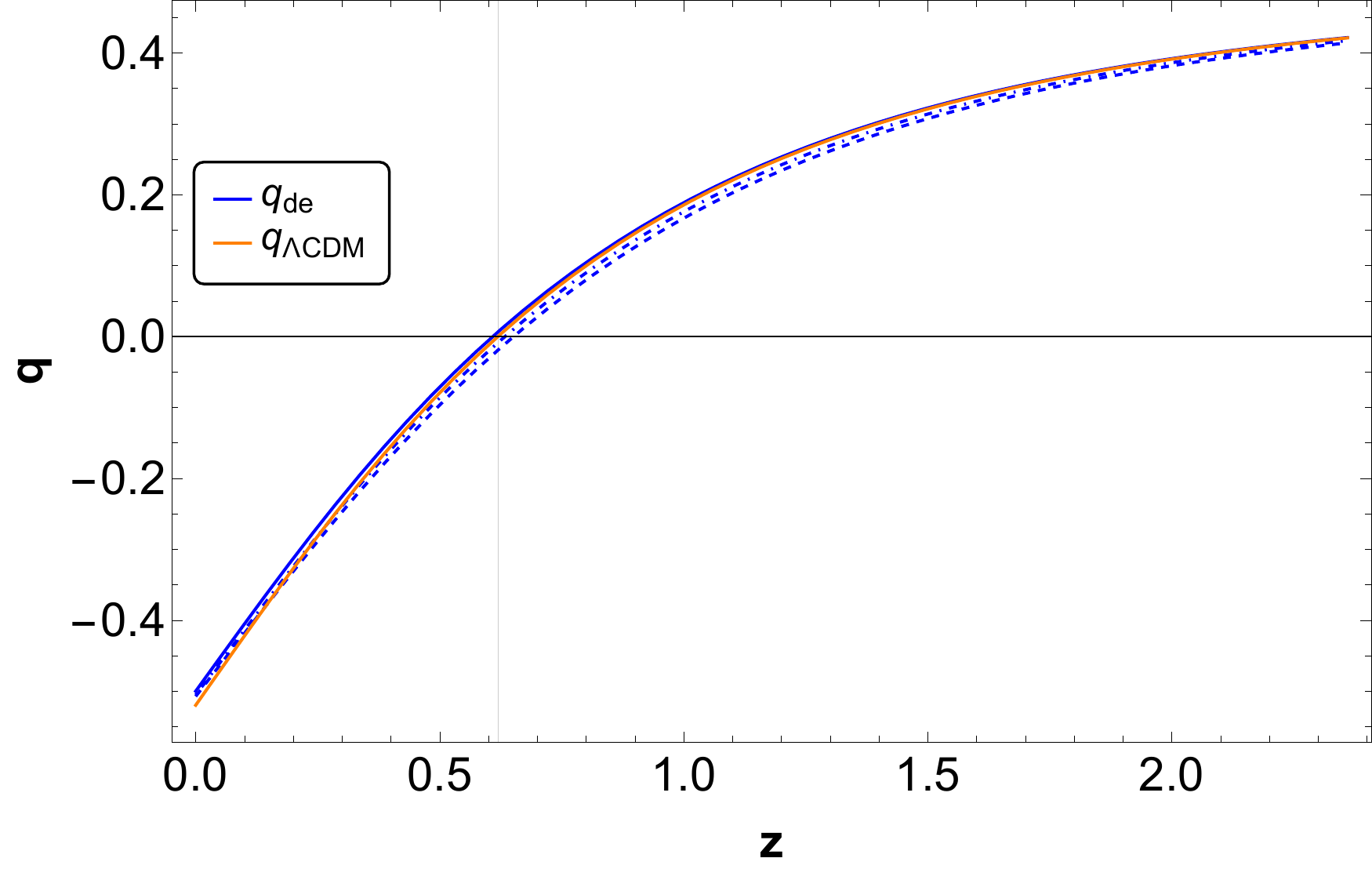}
    \caption{\smaller{We show the evolution of the deceleration parameter $q(z)$, for the same initial conditions used in figure \ref{Figura1}. Where, the transition to dark energy dominance is around $z \approx 0.62$, which is consistent with current observational data.}} 
    \label{Figura10}
  \end{minipage}
\end{figure}


\begin{figure}[!tbp]
  \centering
  \begin{minipage}[b]{0.45\textwidth}
    \includegraphics[width=\textwidth, height=4cm]{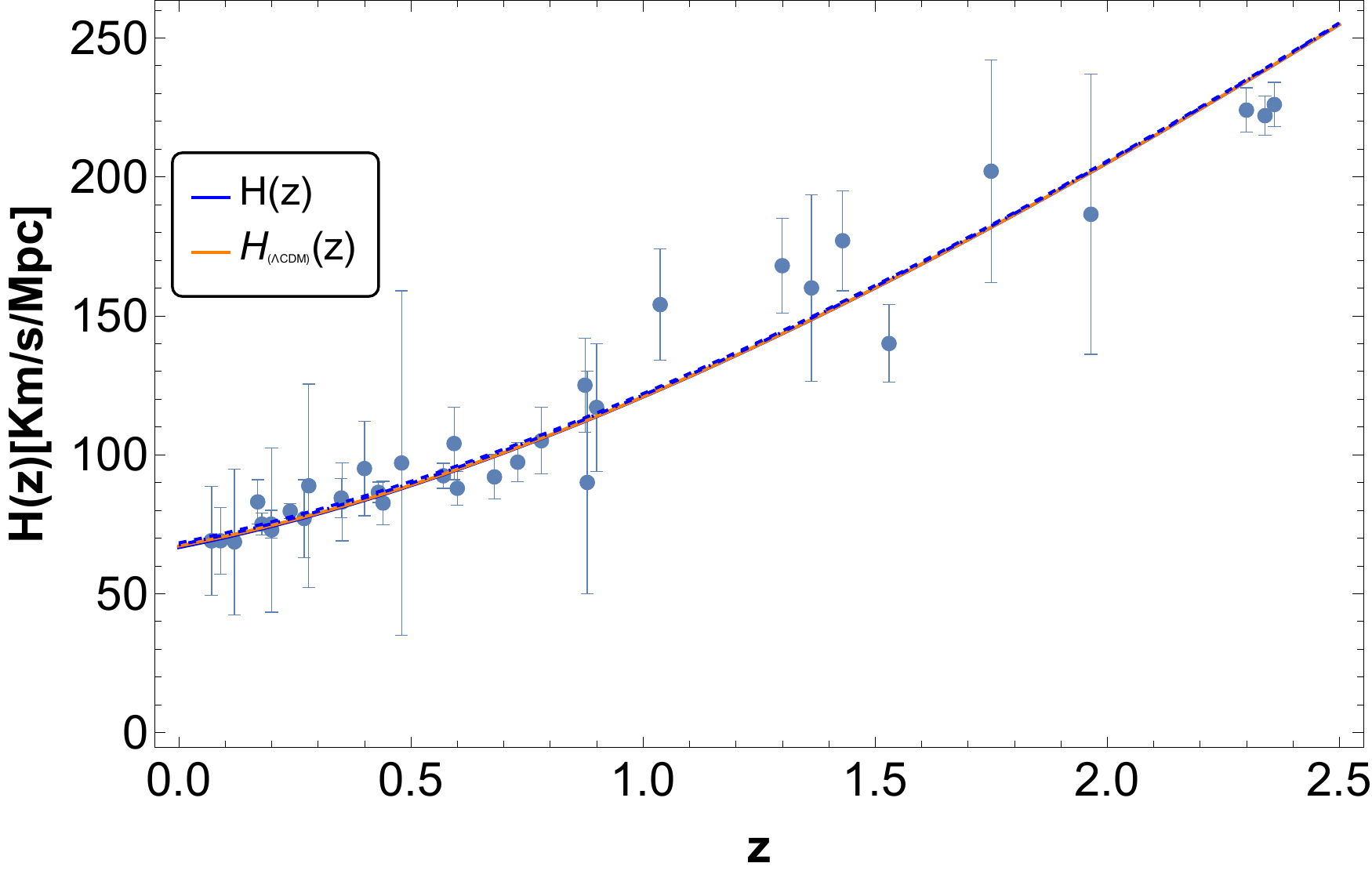}
    \caption{\scriptsize{We present the variation of the Hubble rate, denoted as $H(z)$, in relation to the redshift $z$. The observed data points for the Hubble rate are depicted as blue points, accompanied by their corresponding $1 \sigma$ confidence intervals, as detailed in Table \ref{table:H(z)data}. We utilize identical initial conditions as displayed in FIG \ref{Figura6} to generate the solid, dashed, and dot-dashed blue lines. By comparing the theoretical model predictions represented by these lines with the observed data, we can assess the level of agreement between the empirical measurements of the Hubble rate at different redshifts and the theoretical model.\\}} 
    \label{Figura2H}
  \end{minipage}
  \hfill
  \begin{minipage}[b]{0.45\textwidth}
    \includegraphics[width=\textwidth, height=4cm]{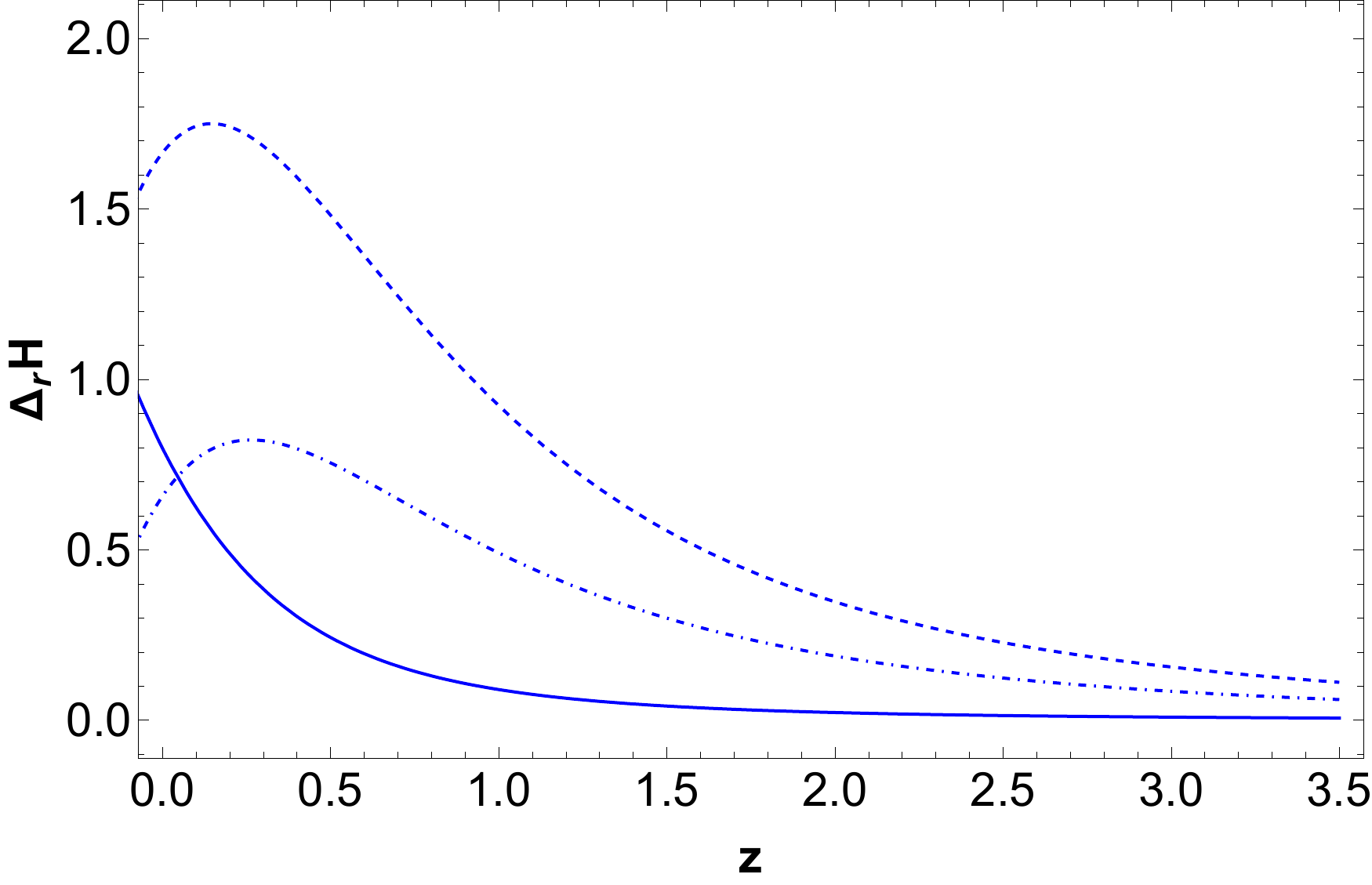}
    \caption{\scriptsize{The figure illustrates the relative difference (expressed as a percentage), denoted as $\Delta_r H(z) = 100 \times \left| H - H_{\Lambda CDM} \right|/H_{\Lambda CDM}$, in comparison to the $\Lambda$CDM model. The solid, dashed, and dot-dashed blue lines are derived from the identical initial conditions presented in FIG. \ref{Figura6}, are utilized in this analysis. By evaluating the relative difference between the Hubble rate $H$ and the Hubble rate of the $\Lambda$CDM model $H_{\Lambda CDM}$, we can examine the deviations from the $\Lambda$CDM model and the associated significance at different redshifts.}} 
    \label{FiguradH2}
  \end{minipage}
\end{figure}


\begin{figure}[!tbp]
	\centering
		\includegraphics[width=0.35\textwidth]{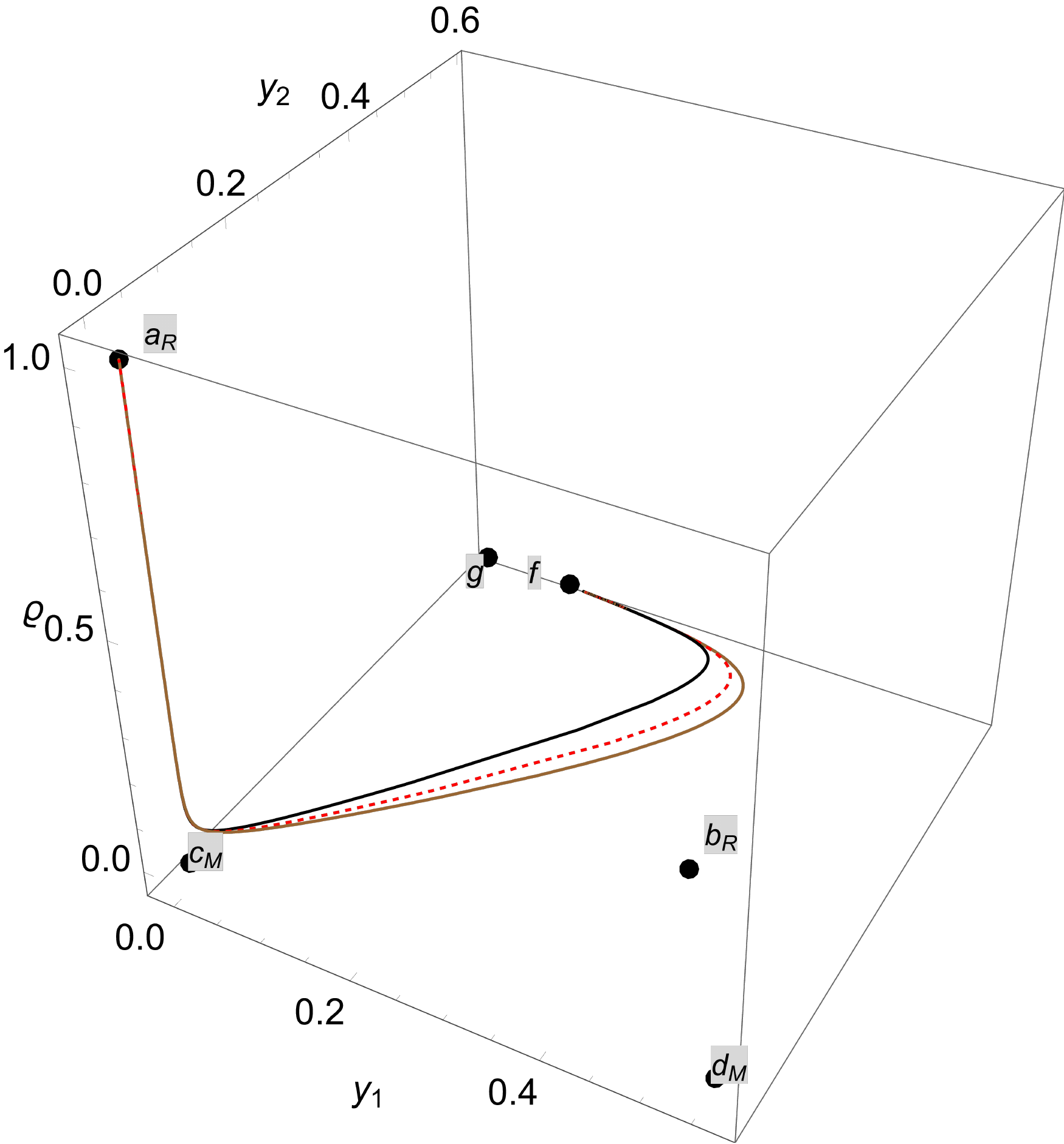}
	\caption{\smaller{Evolution curves in the phase space for $\lambda_1 = 2.2$, $\lambda_2 = 1$ and $\sigma_1 = 1$. Black solid line correspond to initial conditions $x_{1 i} = 10^{-5}$, $x_{2 i} = 10^{-5}$, $x_{3 i} = 10^{-13}$, $y_{1 i} = 3.2 \times 10^{-13}$, $y_{2 i} = 4.5 \times 10^{-13}$, $u_{i} = 4 \times 10^{-13}$ and $\varrho_{i} = 0.99983$. Brown solid line correspond to initial conditions $x_{1 i} = 10^{-5}$, $x_{2 i} = 10^{-5}$, $x_{3 i} = 10^{-13}$, $y_{1 i} = 4.2 \times 10^{-13}$, $y_{2 i} = 4.5 \times 10^{-13}$, $u_{i} = 4 \times 10^{-13}$ and $\varrho_{i} = 0.99983$. Red dot-dashed line correspond to initial conditions $x_{1 i} = 10^{-5}$, $x_{2 i} = 10^{-5}$, $x_{3 i} = 10^{-13}$, $y_{1 i} = 3.8 \times 10^{-13}$, $y_{2 i} = 4.5 \times 10^{-13}$, $u_{i} = 4.0 \times 10^{-13}$ and $\varrho_{i} = 0.99983$.}} 
	\label{Figura9}
\end{figure}


\begin{figure}[!tbp]
  \centering
  \begin{minipage}[b]{0.45\textwidth}
    \includegraphics[width=\textwidth, height=4cm]{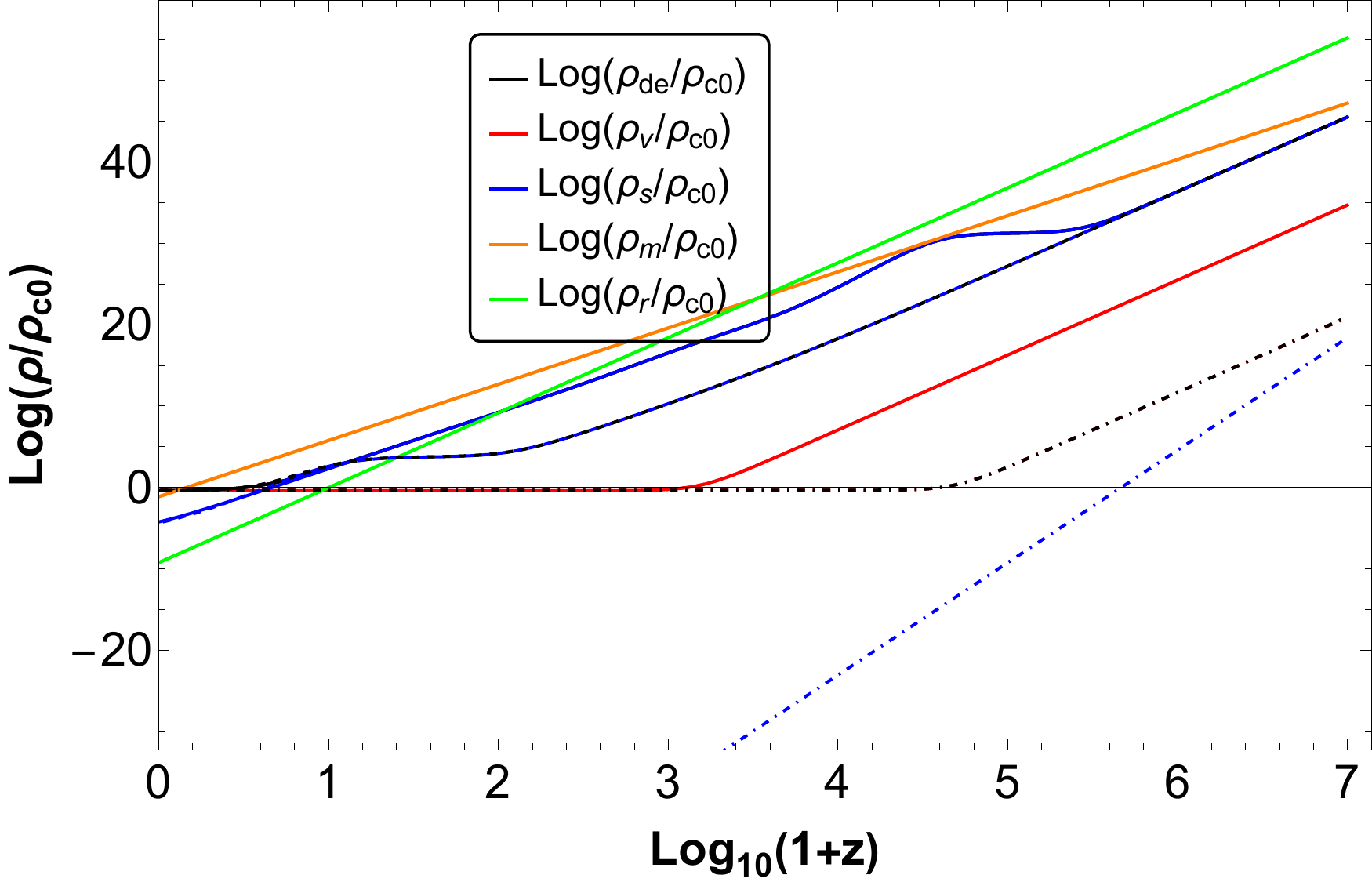}
    \caption{\smaller{We depict the evolution of the energy density of dark energy $\rho_{de}$ (black), scalar field density $\rho_{s}$ (blue), vector field density $\rho_{v}$ (red), dark matter (including baryons) $\rho_m$ (orange) and radiation $\rho_r$ (green) as functions of the redshift z, for $\lambda_1 = 10$, $\lambda_2 = 1$ and $\sigma_1 = 10$. In particular, solid lines correspond to initial conditions $x_{1 i} = 10^{-5}$, $x_{2 i} = 10^{-5}$, $x_{3 i} = 10^{-5}$, $y_{1 i} = 6 \times 10^{-6}$, $y_{2 i} = 4.8 \times 10^{-13}$, $u_{i} = 6 \times 10^{-5}$ and $\varrho_{i} = 0.99983$. Dashed lines correspond to initial conditions $x_{1 i} = 10^{-5}$, $x_{2 i} = 10^{-5}$, $x_{3 i} = 10^{-5}$, $y_{1 i} = 6 \times 10^{-12}$, $y_{2 i} = 4.8 \times 10^{-13}$, $u_{i} = 6 \times 10^{-5}$ and $\varrho_{i} = 0.99983$. Dot-dashed lines correspond to initial conditions $x_{1 i} = 10^{-8}$, $x_{2 i} = 10^{-8}$, $x_{3 i} = 10^{-8}$, $y_{1 i} = 0$, $y_{2 i} = 4.9 \times 10^{-13}$, $u_{i} = 0$ and $\varrho_{i} = 0.99983$. During the scaling radiation era $a_{R}$ we obtained $\Omega_{de}^{(r)}\approx 6.00013 \times 10^{-5}$, while in the case of $b_{R}$ we found {$\Omega_{de}^{(r)}\approx 6.07537 \times 10^{-5}$}. Also, during the scaling matter era $d_{M}$, we got {$\Omega_{de}^{(m)}\approx 1.93572 \times 10^{-2}$} at redshift $z=50$.\\ \\}} 
    \label{Figura11}
  \end{minipage}
  \hfill
  \begin{minipage}[b]{0.45\textwidth}
    \includegraphics[width=\textwidth, height=4cm]{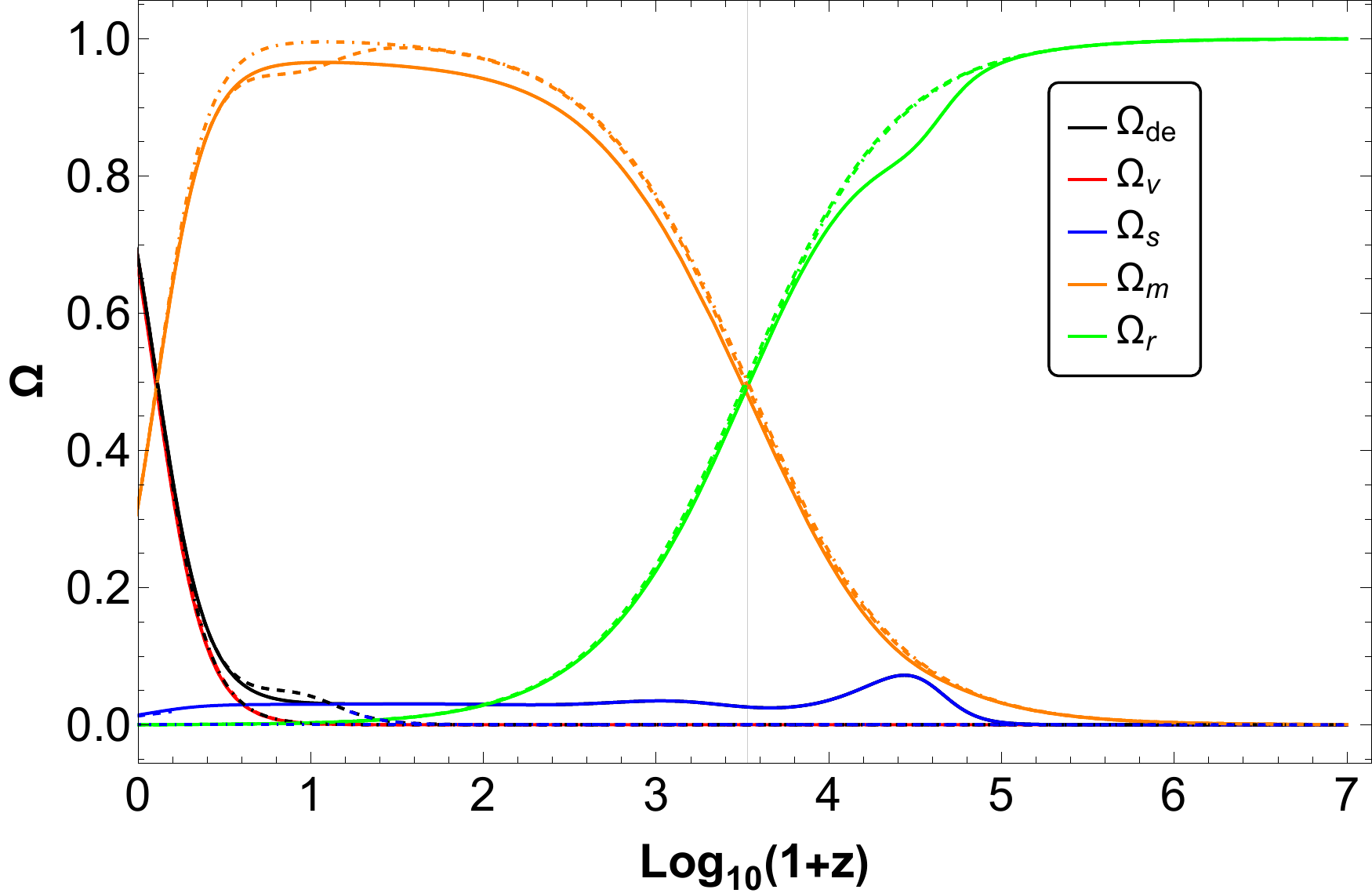}
    \caption{\smaller{The evolution of the fractional energy densities of dark energy (black line), vector field (red line) and  scalar field (blue line), for the same initial conditions used in figure \ref{Figura7}. Also, the time of radiation-matter equality is around $z \approx 3387$.  Finally, at future times, $\Omega_v \approx 0.998365$ and $\Omega_s \approx 1.63459 \times 10^{-3}$.}} 
    \label{Figura12}
  \end{minipage}
\end{figure}

\begin{figure}[!tbp]
  \centering
  \begin{minipage}[b]{0.45\textwidth}
    \includegraphics[width=\textwidth, height=4cm]{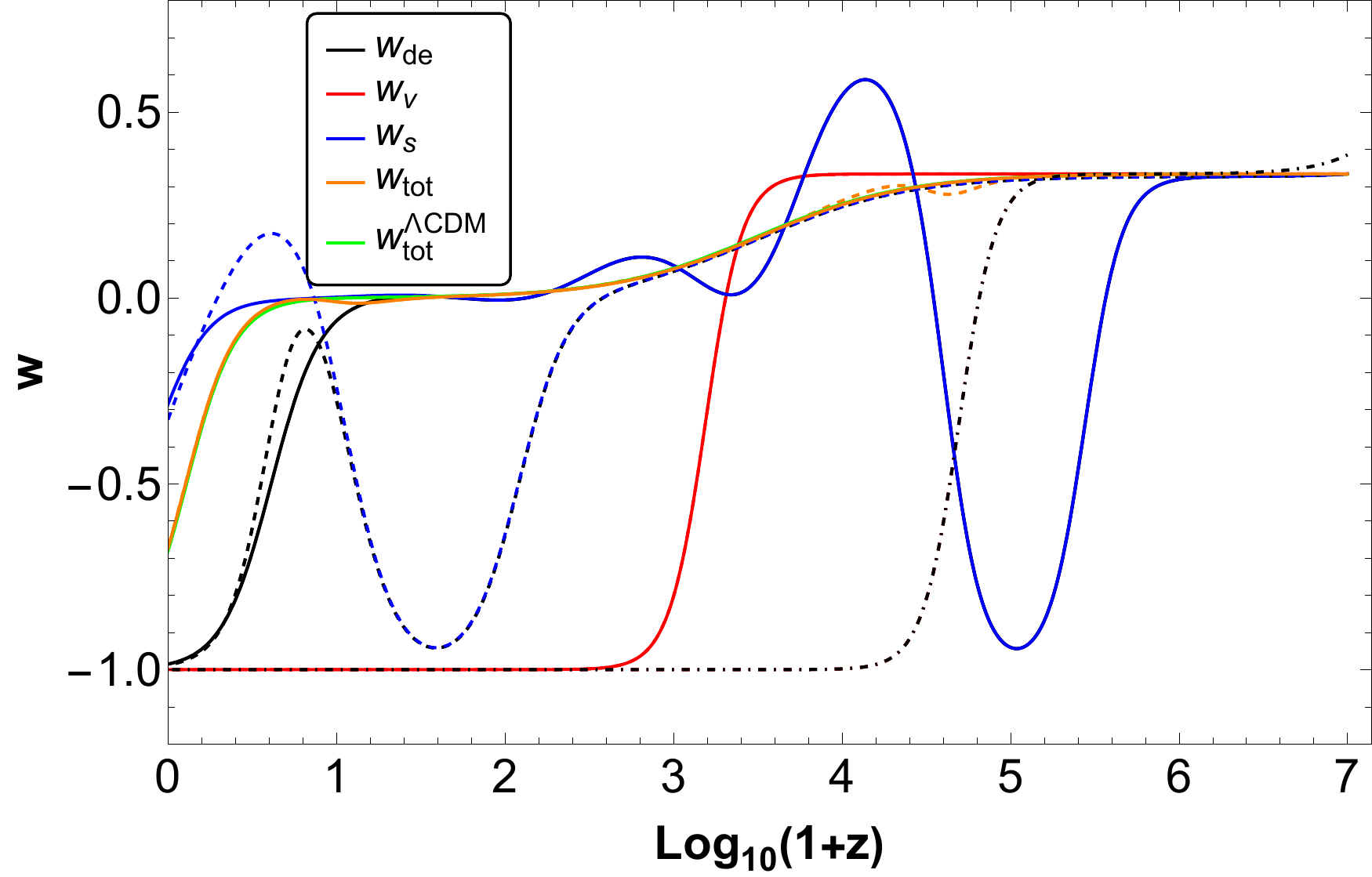}
    \caption{\smaller{We depict the equation of state $w_{tot}$ (orange line), the EOS of dark energy $w_{de}$ (black line), the EOS related to vector field $w_{v}$ (red line), the EOS related to scalar field $w_{s}$ (blue line), and the total EOS of $\Lambda$CDM model (green line) as redshift functions.  Also, we used the same initial conditions of figure \ref{Figura7} to obtain the solid, dashed, and dot-dashed blue lines and red lines. Finally, we obtain the value $w_{de} = -0.997323$ at $z = 0$.\\ \\}} 
    \label{Figura13}
  \end{minipage}
  \hfill
  \begin{minipage}[b]{0.45\textwidth}
    \includegraphics[width=\textwidth, height=4cm]{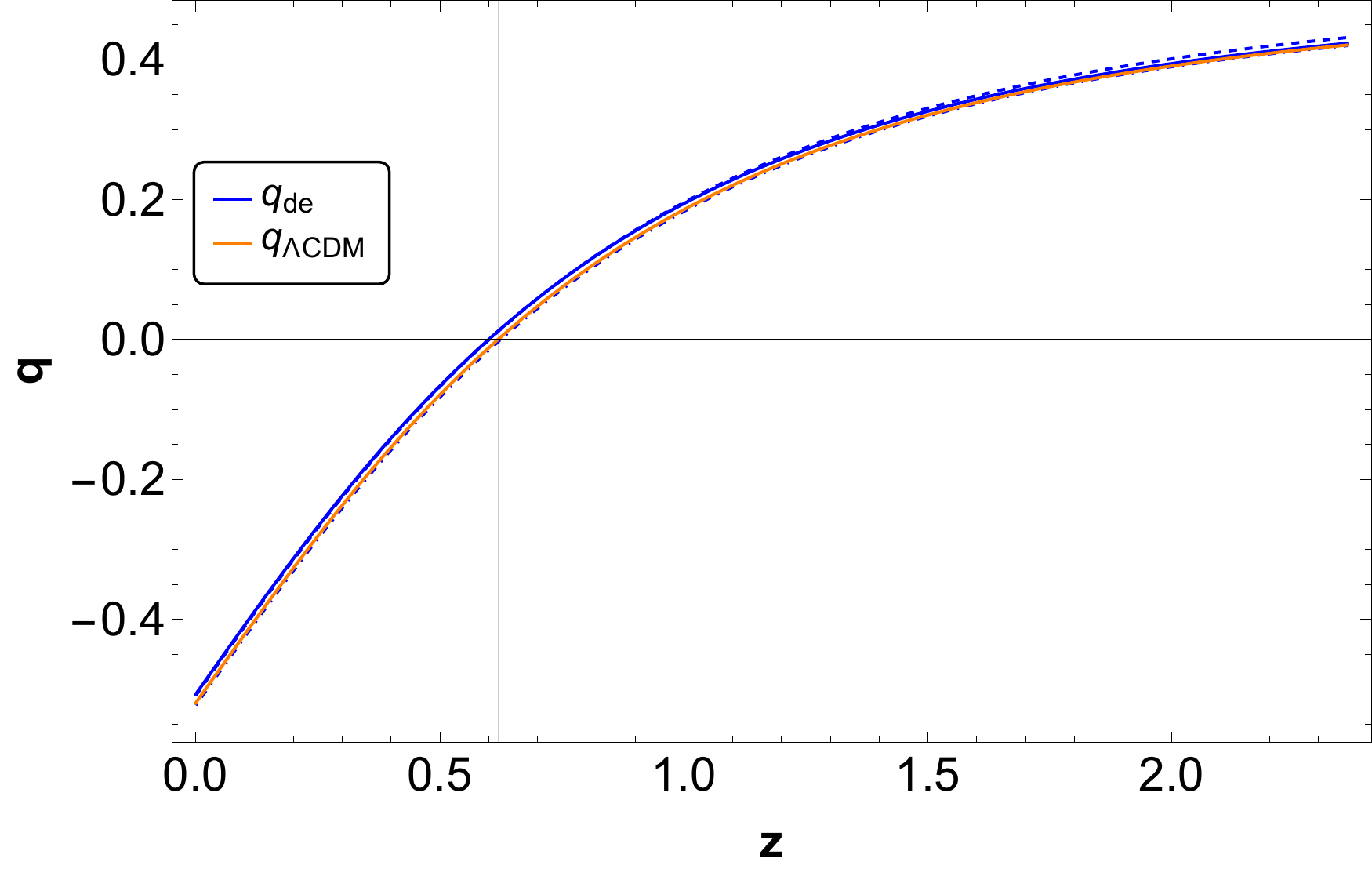}
    \caption{\smaller{We show the evolution of the deceleration parameter $q(z)$, for the same initial conditions used in figure \ref{Figura1}. Where, the transition to dark energy dominance is around $z \approx 0.62$, which is consistent with current observational data.}} 
    \label{Figura15}
  \end{minipage}
\end{figure}


\begin{figure}[!tbp]
  \centering
  \begin{minipage}[b]{0.45\textwidth}
    \includegraphics[width=\textwidth, height=4cm]{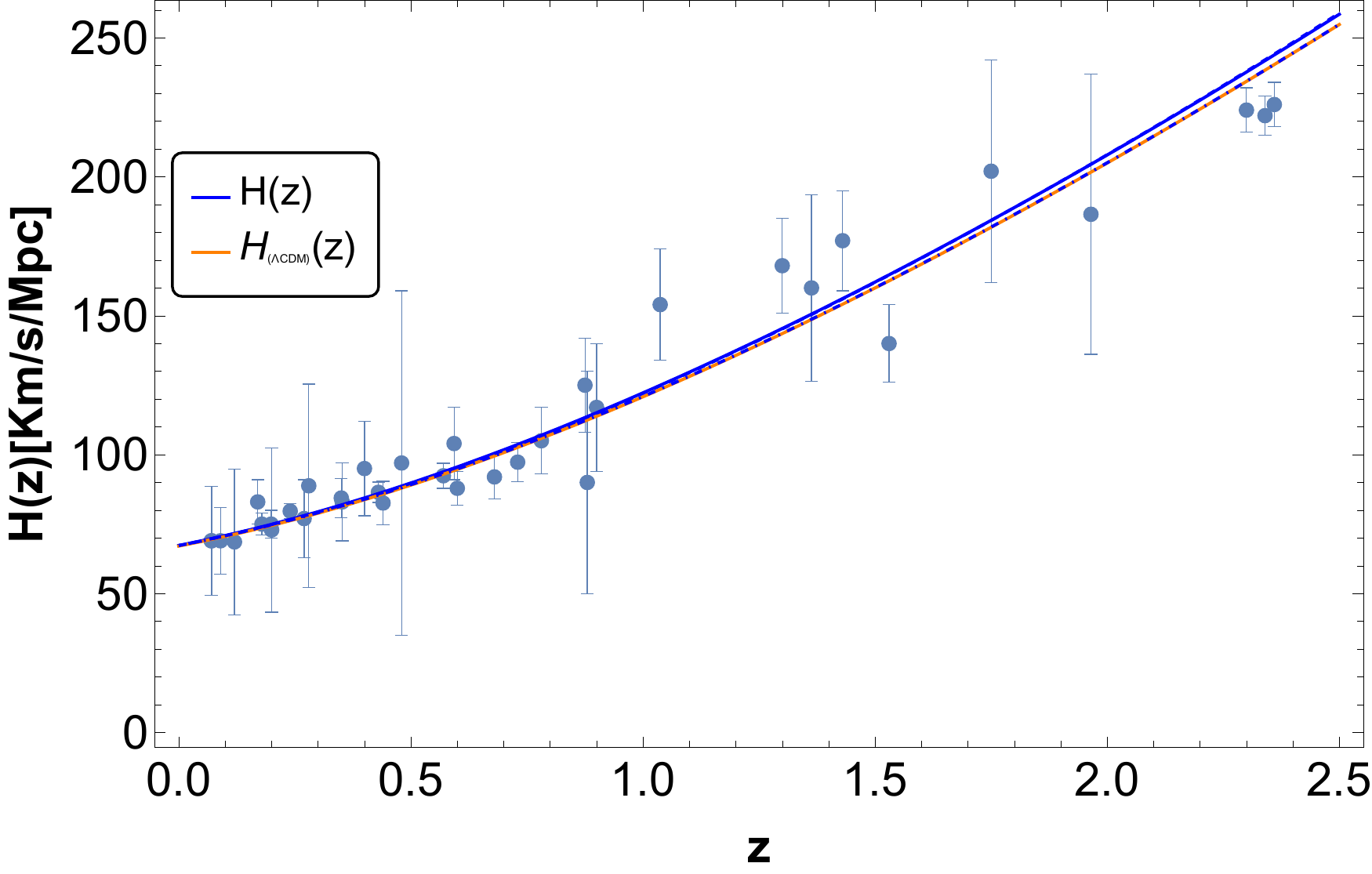}
    \caption{\scriptsize{We present the variation of the Hubble rate, denoted as $H(z)$, in relation to the redshift $z$. The observed data points for the Hubble rate are depicted as blue points, accompanied by their corresponding $1 \sigma$ confidence intervals, as detailed in Table \ref{table:H(z)data}. We utilize identical initial conditions as displayed in FIG \ref{Figura11} to generate the solid, dashed, and dot-dashed blue lines. By comparing the theoretical model predictions represented by these lines with the observed data, we can assess the level of agreement between the empirical measurements of the Hubble rate at different redshifts and the theoretical model.\\}} 
    \label{Figura3H}
  \end{minipage}
  \hfill
  \begin{minipage}[b]{0.45\textwidth}
    \includegraphics[width=\textwidth, height=4cm]{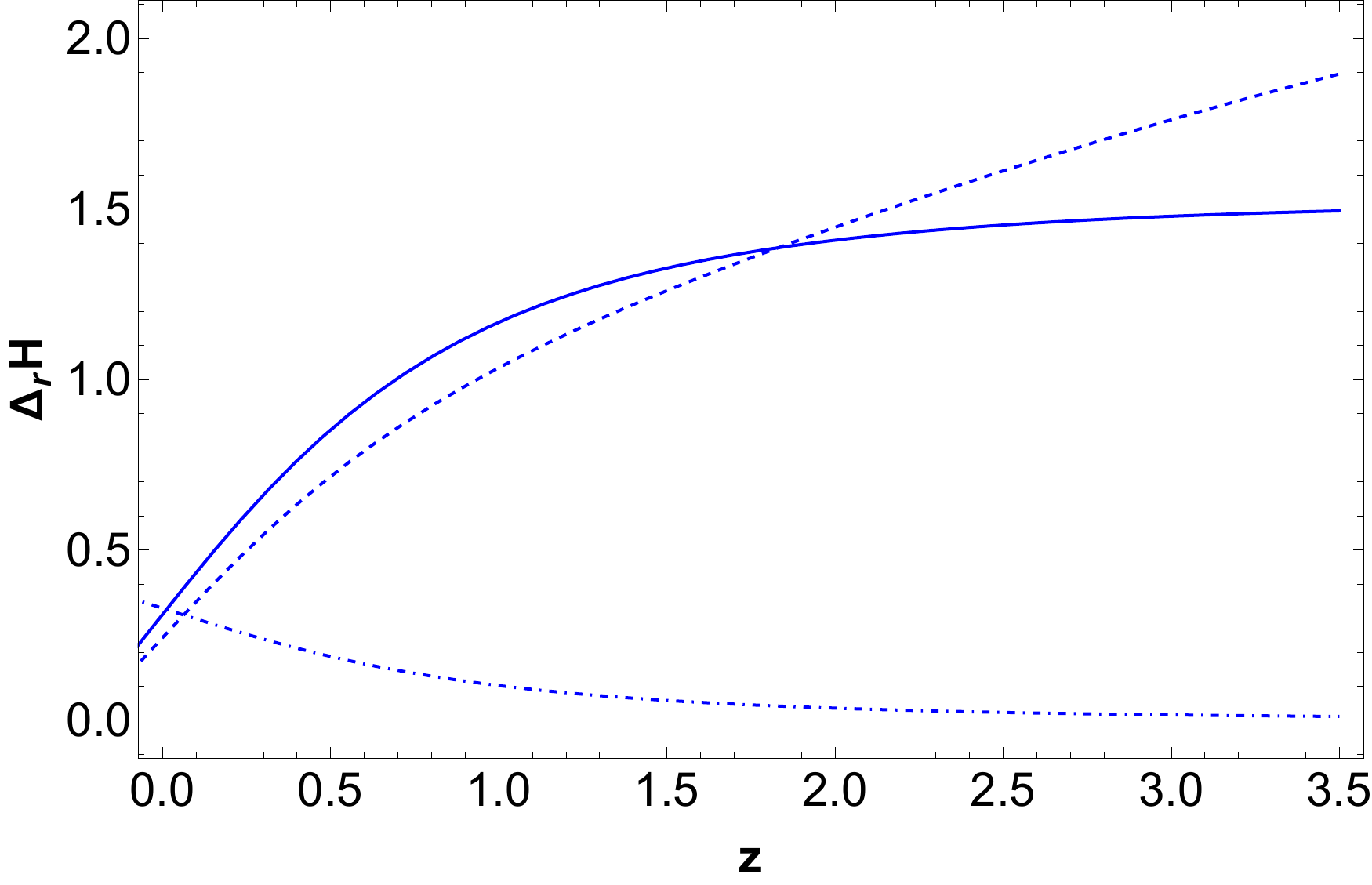}
    \caption{\scriptsize{The figure illustrates the relative difference (expressed as a percentage), denoted as $\Delta_r H(z) = 100 \times \left| H - H_{\Lambda CDM} \right|/H_{\Lambda CDM}$, in comparison to the $\Lambda$CDM model. The solid, dashed, and dot-dashed blue lines are derived from the identical initial conditions presented in FIG. \ref{Figura11}, are utilized in this analysis. By evaluating the relative difference between the Hubble rate $H$ and the Hubble rate of the $\Lambda$CDM model $H_{\Lambda CDM}$, we can examine the deviations from the $\Lambda$CDM model and the associated significance at different redshifts.}} 
    \label{FiguradH3}
  \end{minipage}
\end{figure}


\begin{figure}[!tbp]
	\centering
		\includegraphics[width=0.35\textwidth]{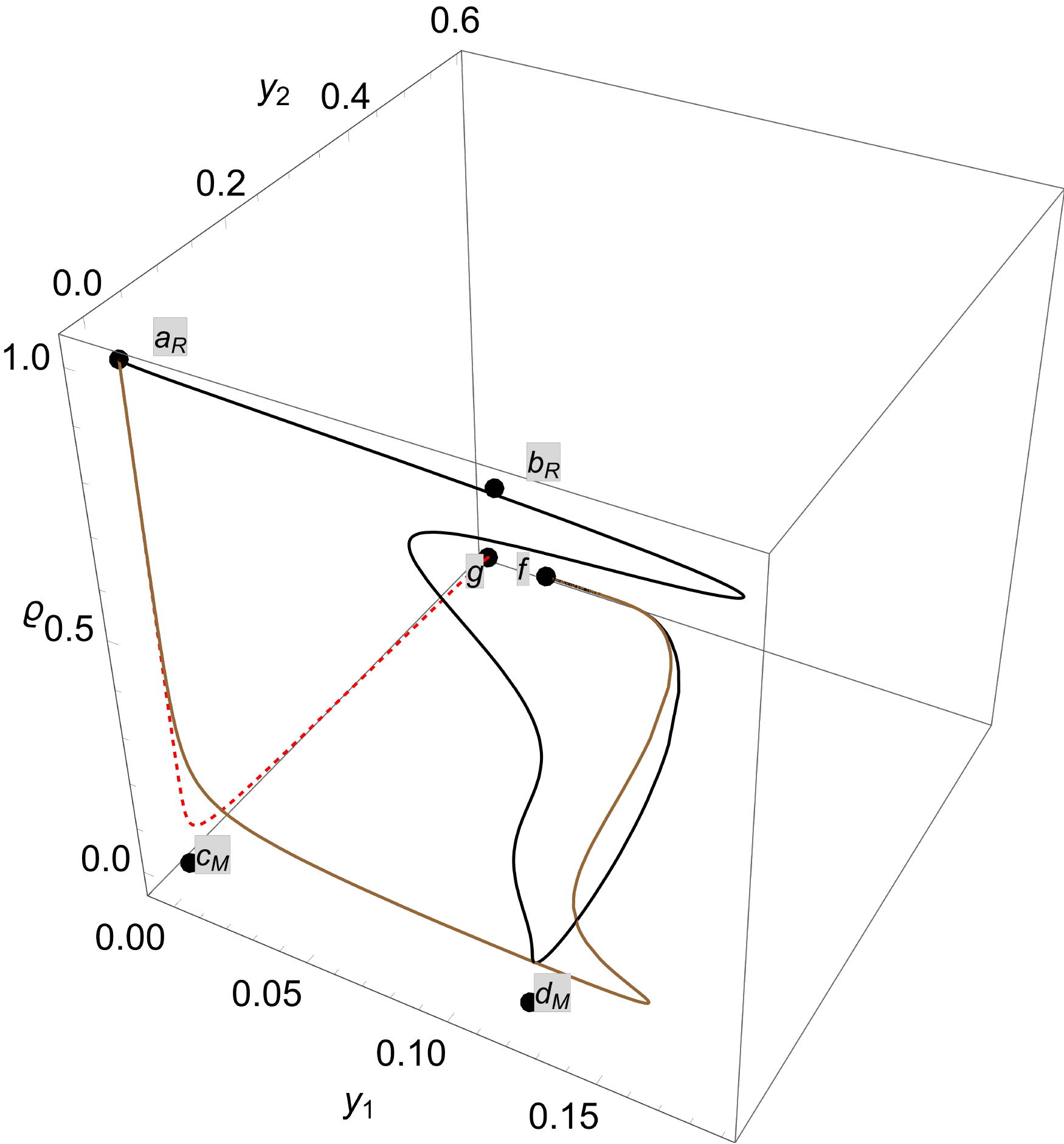}
	\caption{\smaller{ Evolution curves in the phase space for $\lambda_1 = 10$, $\lambda_2 = 1$ and $\sigma_1 = 10$. Black solid line correspond to initial conditions $x_{1 i} = 10^{-5}$, $x_{2 i} = 10^{-5}$, $x_{3 i} = 10^{-5}$, $y_{1 i} = 6 \times 10^{-6}$, $y_{2 i} = 4.8 \times 10^{-13}$, $u_{i} = 6 \times 10^{-5}$ and $\varrho_{i} = 0.99983$. Brown solid line correspond to initial conditions $x_{1 i} = 10^{-5}$, $x_{2 i} = 10^{-5}$, $x_{3 i} = 10^{-5}$, $y_{1 i} = 6 \times 10^{-12}$, $y_{2 i} = 4.8 \times 10^{-13}$, $u_{i} = 6 \times 10^{-5}$ and $\varrho_{i} = 0.99983$. Red dot-dashed line correspond to initial conditions $x_{1 i} = 10^{-8}$, $x_{2 i} = 10^{-8}$, $x_{3 i} = 10^{-8}$, $y_{1 i} = 0$, $y_{2 i} = 4.9 \times 10^{-13}$, $u_{i} = 0$ and $\varrho_{i} = 0.99983$.}} 
	\label{Figura14}
\end{figure}


\section{Numerical Results}\label{Num_Res}

We found five critical points that allow an accelerated expansion, $f, g, h, i$ and $j$. And we have verified numerically that the points $f$, $g$, $h$, and $j$ are attractors. However, the critical point $i$, which is related to the ordinary exponential quintessence model \cite{Copeland:2006wr,amendola2010dark} is no longer a stable point, Eq. \eqref{cp_i}.

In FIG. \ref{Figura4}, we show the evolution of the system in the phase space through the trajectories $a_{R}\rightarrow c_{M}\rightarrow  f$, $a_{R}\rightarrow d_{M}\rightarrow  f$ and $b_{R}\rightarrow d_{M}\rightarrow  f$. In FIG. \ref{Figura1} it is depicted the evolution of the energy densities of dark energy, dark matter and radiation, along with the vector and the scalar components of dark energy. Thus, one can verify that there exists for these trajectories a scaling regime of the fields during the radiation and matter eras. Furthermore, in FIG. \ref{Figura2} we plot the behavior of the corresponding fractional energy densities for the same initial conditions that in FIG \ref{Figura1}, showing that the nature of the final attractor is purely vectorial. At the present time, redshift $z=0$, we found $\Omega_{de}^{(0)}\approx 0.68$ and dark matter $\Omega_{m}^{(0)}\approx 0.32$, which is consistent with observations.  

In FIG. \ref{Figura3} we show the cosmic evolution of the total EOS, the EOS of dark energy, and the EOS parameter of the vectorial and scalar components of dark energy. To compare with the current concordance model, we also added the evolution curve of the EOS of the $\Lambda$CDM model. For the EOS parameter of dark energy at $z=0$ we obtained  $w_{de}(z=0)\approx -0.999516$, which is in agreement with the observational constraint $w_{de}^{(0)}=-1.028\pm 0.032$ \cite{Aghanim:2018eyx}. From FIG. \ref{Figura2} one can check that the time of radiation-matter equality is around $z\approx 3387$. Moreover, in FIG. \ref{Figura5} we depict the deceleration parameter for the same initial conditions. This plot shows that the transition to the accelerated phase occurs at $z\approx 0.62$, which is very close to the $\Lambda$CDM value. 

It is important to mention that for the scaling radiation/matter regimes, we have respected the constraints coming from both the Physics of Big Bang Nucleosynthesis (BBN), $\Omega_{de}^{(r)}<0.045$, \cite{Bean:2001wt}, and CMB measurements from Planck, $\Omega_{de}^{(m)}<0.02$ ($95 \%$ CL) at redshift $z= 50$ \cite{Ade:2015rim}. For instance, in FIG \ref{Figura1},  during the scaling radiation era $a_{R}$ we obtained $\Omega_{de}^{(r)}\approx 9.91736 \times 10^{-4}$, while in the case of $b_{R}$ we found {$\Omega_{de}^{(r)}\approx 1.66852 \times 10^{-3}$}. Also, during the scaling matter era $d_{M}$, we got {$\Omega_{de}^{(m)}\approx 1.00912 \times 10^{-3}$} at redshift $z=50$.

In FIGS. \ref{Figura9} and \ref{Figura14} we depict the evolution curves in the phase space for the transitions   $a_{R}\rightarrow c_{M}\rightarrow  f$, $a_{R}\rightarrow c_{M}\rightarrow  g$,  $a_{R}\rightarrow c_{M}\rightarrow d_{M}\rightarrow  f$, and  $a_{R}\rightarrow b_{R}\rightarrow d_{R}\rightarrow  f$,
for another set of parameter values and initial conditions. The corresponding behavior of the energy densities of dark energy, dark matter and radiation are shown in  FIGS. \ref{Figura6}, \ref{Figura7}, \ref{Figura11} and \ref{Figura12} . Moreover, the cosmic evolution of the EOS parameter of dark energy and the total EOS are shown in FIGS \ref{Figura8} and \ref{Figura13}. In these figures, we also depicted the behavior of the vector and scalar fields.  Finally, the evolution of the deceleration parameter is presented in FIGS. \ref{Figura10} and \ref{Figura15}. 

{
We choose parameters $\lambda_1$, $\lambda_2$, and $\sigma$ based on the stability conditions of critical points and observational constraints. Specifically, we numerically solve the dynamical system (41) using three sets of values for the parameters $\lambda_1$, $\lambda_2$, and $\sigma$, as well as initial conditions (values given in Fig. 1, Fig. 8, and Fig. 15). These values are selected to ensure the stability of points $f$ and $g$ (Fig. 7, Fig. 14, and Fig. 21) and to be consistent with the observational values: $\Omega_{DE} \approx 0.68$ (Fig. 2, Fig. 9, and Fig. 16), $w^{(0)}_{de} = -1.028 \pm 0.032$ (Fig. 3, Fig. 10, and Fig. 17), radiation-matter equality around $z \approx 3387$ (Fig. 2, Fig. 9, and Fig. 16), and transition to dark energy dominance around $z \approx 0.62$ (Fig. 4, Fig. 11, and Fig. 18). 
Even more, our numerical solutions (Fig. 2, Fig. 9, and Fig. 16) are consistent with constraints arising from Big Bang Nucleosynthesis (BBN) and measurements obtained from the Cosmic Microwave Background (CMB) by the Planck satellite. That is to say, according to BBN, $\Omega^{(r)}_{de}$ is limited to $\Omega^{(r)}_{de}<0.045$ ($2.\sigma$ C.L) \cite{Bean:2001wt}, while CMB measurements at a redshift of approximately $z \approx 50$ from Planck indicate $\Omega^{(m)}_{de}<0.02$ ($2.\sigma$ C.L) \cite{Ade:2015rim}. Finally, it is important to note that the next step involves identifying the best-fit parameters of $\lambda_1$, $\lambda_2$, and $\sigma$ and determining their corresponding confidence intervals through statistical analysis. However, conducting this analysis exceeds the scope of this paper.
}
But, with these previous results, we have proved that the model satisfies the preliminary requirements to be considered viable \cite{Aghanim:2018eyx}. 
\section{Hubble parameter $H(z)$}
{
To analyze the behavior of the Hubble rate and its confidence interval, we will utilize a set of $39$ data points obtained from previous studies \cite{Farooq:2016zwm,Ryan:2018aif}. These data points, expressed in terms of redshift $z$, are presented in Table \ref{table:H(z)data}.
Moreover, to facilitate further comparisons, we employ the $\Lambda$CDM model, which provides a Hubble rate varying with redshift according to the following expression:
\be
H_{\Lambda CDM}(z)=H_{0}\sqrt{\Omega_{de}^{(0)}+\Omega_{m}^{(0)}(1+z)^3+\Omega_{r}^{(0)}(1+z)^4} .
\ee
Figures \ref{Figura1H}, \ref{Figura2H} and \ref{Figura3H} illustrate the evolution of the Hubble rate as a function of redshift for our model. The parameter values and initial conditions utilized for Figures \ref{Figura1}, \ref{Figura6} and \ref{Figura11} are consistent with those employed for the aforementioned plots. We compare these results with the standard outcomes obtained from the $\Lambda$CDM model and the current observational data on the Hubble rate $H(z)$. Consequently, we conclude that our findings align with observational $H(z)$ data. To further evaluate the consistency between our model and $\Lambda$CDM, we examine the relative difference, denoted as $\Delta_r H(z)$, between the results obtained from our model and $\Lambda$CDM. Figures \ref{FiguradH1}, \ref{FiguradH2} and \ref{FiguradH3} provide a visualization of this comparison, indicating that our results closely resemble those of $\Lambda$CDM.
}

\section{Discussions and final Remarks}\label{conclusion_f}
We investigated the cosmological dynamics of dark energy in a scalar-vector-torsion model \cite{Armendariz-Picon:2004say}, where the vector part is described by the cosmic triad, the scalar field is of quintessence type, and the torsion is associated to the Weitzenb\"{o}ck connection of teleparallel gravity  \cite{Einstein,TranslationEinstein,Early-papers1,Early-papers2,Early-papers3,Early-papers4,Early-papers5,Early-papers6}. We performed an exhaustive phase analysis looking for the critical points of the autonomous system and their stability properties \cite{Copeland:2006wr}. 

In order to find the set of cosmological equations that govern the dynamics of the fields, we have assumed the FLRW background in the presence of radiation and cold dark matter fluids. We obtained the effective energy and pressure densities of dark energy and identified contributions from both the vector and scalar fields. In this way, 
we derived the autonomous system from this set of cosmological equations, and the corresponding critical points are shown in Table \ref{table1}. Furthermore, in Table \ref{table2}, we presented for each fixed point the results obtained for the relevant cosmological parameters. So, by numerically integrating the autonomous system we have corroborated all our analytical results. Particularly, we have shown that the dark energy model studied here, which is based on a scalar-vector-torsion gravity theory, is cosmologically viable and capable of successfully reproducing the thermal history of the universe. We found new attractor points, which are dark energy-dominated solutions with accelerated expansion. 
For these critical points, the fractional energy density of dark energy is composed 
of contributions coming from both the scalar and vector field densities. Thus, the dynamics of dark energy is driven by the two fields. The vector field behaves close to the cosmological constant, while the scalar field remains in the quintessence regime. Nevertheless, at late times we found that the energy density of the scalar field is always subdominant, and therefore the nature of dark energy could be mainly vectorial.  

We also obtained new scaling solutions in which the scalar and vector field densities scale in the same way as the radiation and matter background fluids. After performing the stability analysis, we found that they are unstable fixed points without accelerated expansion. Although the vector field density only can scale as radiation and not matter, the scalar field can mimic both \cite{amendola2010dark}. Also, it is important to note that scaling solutions are attractive once they can naturally incorporate early dark energy. They provide us with a mechanism to alleviate the energy scale problem of the $\Lambda$CDM model, which is related to the large energy gap that exists between the current critical energy density and the typical energy scales of particle physics \cite{Albuquerque:2018ymr,Ohashi:2009xw}. 

{We found that the effective dark energy, comprising the energy densities of the scalar and vector fields, can exhibit scaling behavior in the early universe. This implies the existence of a small portion of dark energy contributing to the dynamics of the universe during the radiation and matter eras. It is important to note that both the vector and the scalar can drive this scaling behavior during the radiation era, while in the matter era, only the scalar field can provide this effect. This is a natural outcome, as an oscillating field can indeed act like an effective fluid with a constant equation of state, which has radiation-like or matter-like behavior depending on the potential \cite{MukhanovBook}.  Moreover, the dynamics of the vector field resemble that of a Maxwell-type field \cite{Koivisto:2008xf}.} 

{Since the scaling solutions lead to the presence of small amounts of dark energy during the radiation and matter era, significant physical consequences can emerge \cite{Albuquerque:2018ymr,Ohashi:2009xw}. Specifically, it alters the Hubble rate during this period, thereby affecting the theoretical predictions for the abundances of primordial light elements \cite{Ferreira:1997hj,Bean:2001wt}. Moreover, a scaling field significantly impacts the shape of the spectrum of Cosmic Microwave Background (CMB) anisotropies \cite{Bean:2001wt}. This scaling field also influences the formation of large-scale structures in the universe, impacting the growth of cosmic structures such as galaxies and clusters of galaxies  \cite{Amendola:1999er}}. In this latter case, the effects on the matter power spectrum arise from the reduction of the matter fluctuation variance $\sigma_{8}$, due to the suppressed growth rate of matter perturbations \cite{Amendola:1999er,Amendola:1999dr,Davari:2019tni}. {Utilizing the latest observational data from Big Bang Nucleosynthesis and observed abundances of primordial nuclides, the authors in Ref. \cite{Bean:2001wt} established a strong constraint $\Omega_{de}<0.045$ at $2.\sigma$ C.L. for the permissible amount of dark energy during the radiation era. Additionally, based on CMB measurements \cite{Ade:2015rim}, the constraint for the field energy density during the matter scaling epoch is $\Omega_{de}<0.02$ ($2.\sigma$ C.L) at redshift $z \approx 50$. In our study, we used these values to constrain the model parameters when numerically solving our cosmological equations. Although this approach is valid and sufficient from a practical standpoint, for greater precision, we could further investigate the dynamics of the field density during early times to better predict the primordial abundances and the evolution of the cosmological perturbations. However, these additional studies are beyond the scope of the current work and will be the focus of future projects.}

Finally, in order to decide whether or not the present model is a good candidate to describe nature,  more studies are required. Particularly, these studies should include a detailed observational confrontation, using data from Type Ia Supernovae (SNIa), Baryon Acoustic Oscillations (BAO) and Cosmic Microwave Background (CMB) observations. Also, a detailed perturbation analysis to investigate the matter clustering behavior, and the $\sigma_{8}$ tension, would be essential \cite{Abdalla:2022yfr,DiValentino:2020zio,DiValentino:2020vvd,Heisenberg:2022lob}. These interesting and necessary studies lie beyond the scope of the present work and are left for future projects.\\
\\
\textbf{Data availability:} There are no new data associated with this article.\\
\begin{acknowledgments}
M. Gonzalez-Espinoza acknowledges the financial support of FONDECYT de Postdoctorado, N° 3230801. Y. Leyva acknowledges Dirección de Investigación, Postgrado y Transferencia Tecnológica de la
Universidad de Tarapacá for financial support through Proyecto 
``Fortalecimiento Grupos de Investigación UTA Código N° 4732-23". G. Otalora and J. Saavedra acknowledge the financial support of Fondecyt Grant 1220065. \\
\end{acknowledgments}


\bibliography{bio} 

\hfill \break

\newpage

\begin{appendix}
\section{Eigenvalues of points f and g} \label{eigenfg}
Point $f$ has the eigenvalues
    \begin{eqnarray}
        &\mu_1 = 0, \ \ \ \ \mu_2 = -2, \ \ \ \ \mu_3 = -3, \nonumber\\
         & \mu_4 = \frac{1}{2} \left(-\sqrt{24 \lambda _2 y_{2 c}^2+1}-3\right), \nonumber\\
         &\mu_5 = \frac{1}{2} \left(\sqrt{24 \lambda _2 y_{2 c}^2+1}-3\right), \nonumber\\
         &\mu_6 = \dfrac{-\sqrt{3} \sqrt{A}-9 \lambda _1 y_{2 c}^2-3 \sigma _1}{2 \left(3 \lambda _1 y_{2 c}^2+\sigma _1\right)} \nonumber\\
         &\mu_7 = \dfrac{\sqrt{3} \sqrt{A}-9 \lambda _1 y_{2 c}^2-3 \sigma _1}{2 \left(3 \lambda _1 y_{2 c}^2+\sigma _1\right)} ,\label{cp_f}
    \end{eqnarray}
    where,
    \begin{eqnarray*}
        A&=& 108 \lambda _1^3 \sigma _1 y_{2 c}^6-36 \lambda _1^3 \sigma _1 y_{2 c}^4-12 \lambda _1 \sigma _1^3 y_{2 c}^2+18 \lambda _1 \sigma _1 y_{2 c}^2 \nonumber\\
        &+&27 \lambda _1^2 y_{2 c}^4+4 \lambda _1 \sigma _1^3+3 \sigma _1^2 .
    \end{eqnarray*}

Point $g$ has the eigenvalues
    \begin{eqnarray}
        &\mu_1 = 0, \ \ \ \ \mu_2 = -2, \ \ \ \ \mu_3 = -3, \nonumber\\
         & \mu_4 = \dfrac{1}{2} \left(-\sqrt{\dfrac{A-6 \sqrt{B}}{\lambda _2 \left(3 \lambda _1 \lambda _2 x_{3 c}^2+\lambda _2 \sigma _1+\lambda _1\right)}}-3\right), \nonumber\\
         &\mu_5 = \dfrac{1}{2} \left(\sqrt{\dfrac{A-6 \sqrt{B}}{\lambda _2 \left(3 \lambda _1 \lambda _2 x_{3 c}^2+\lambda _2 \sigma _1+\lambda _1\right)}}-3\right), \nonumber\\
         &\mu_6 = \dfrac{1}{2} \left(-\sqrt{\dfrac{A+6 \sqrt{B}}{\lambda _2 \left(3 \lambda _1 \lambda _2 x_{3 c}^2+\lambda _2 \sigma _1+\lambda _1\right)}}-3\right), \nonumber\\
         &\mu_7 = \dfrac{1}{2} \left(\sqrt{\dfrac{A+6 \sqrt{B}}{\lambda _2 \left(3 \lambda _1 \lambda _2 x_{3 c}^2+\lambda _2 \sigma _1+\lambda _1\right)}}-3\right) ,\label{cp_g}
    \end{eqnarray}
    where,
    \begin{eqnarray*}
        A&=& 6 \lambda _1^2 \sigma _1 \left(3 \lambda _2 x_{3 c}^2+1\right) \left(\lambda _2 \left(3 x_{3 c}^2-1\right)+1\right) \nonumber\\
        &+&3 \lambda _2 \lambda _1 \left(2 \sigma _1^2 \left(-3 \lambda _2 x_{3 c}^2+\lambda _2-1\right)+\lambda _2 x_{3 c}^2+3\right) \nonumber\\
        &+&3 \lambda _2^2 \sigma _1 \left(8 x_{3 c}^2 \left(\lambda _2 \left(6 x_{3 c}^2-2\right)+1\right)+3\right) ,
    \end{eqnarray*}
    and,
    \begin{eqnarray*}
        B&=& 16 \lambda _1 \lambda _2^2 \sigma _1 x_{3 c}^2 \left(\lambda _2 \left(3 x_{3 c}^2-1\right)+1\right) \left(3 \lambda _1 \lambda _2 x_{3 c}^2+\lambda _2 \sigma _1+\lambda _1\right) \nonumber\\
        &&\left(\lambda _2 \sigma _1 \left(6 x_{3 c}^2-2\right)+\lambda _1+\sigma _1\right)+\Bigl(3 \lambda _2^2 \left(3 \lambda _1^2+8 \lambda _2\right) \sigma _1 x_{3 c}^4 \nonumber\\
        &&-\lambda _2 x_{3 c}^2 \left(3 \left(\lambda _2-2\right) \lambda _1^2 \sigma _1+\lambda _2 \lambda _1 \left(3 \sigma _1^2+4\right)+4 \lambda _2 \left(2 \lambda _2-1\right) \sigma _1\right) \nonumber\\
        &&+\lambda _1 \left(\lambda _2-1\right) \sigma _1 \left(\lambda _2 \sigma _1-\lambda _1\right)\Bigr)^2 .
    \end{eqnarray*}

\hfill \break

\newpage

\section{Hubble's parameter data}
Within this appendix, we provide the dataset comprising Hubble's parameter values for the redshift range of $0.01 < z < 2.360$:
\begin{table}[b]
\caption{Hubble's parameter vs. redshift \& scale factor.}

\label{table:H(z)data}
\renewcommand{\tabcolsep}{0.7pc} 
\renewcommand{\arraystretch}{0.7} 
\begin{tabular}{@{}lllll}
\hline \hline
  $\;\; z$    &  $ H(z) \;$ ($\frac{km/s}{\text{Mpc}}$ ) &  Ref. \\
\hline
$0.07$      & $ \; \qquad 69     \pm 19.6 $      &   \cite{zhang2014} \\
$0.09$      & $ \; \qquad 69     \pm 12 $      & \cite{simon2005} \\
$0.100$     & $ \; \qquad 69     \pm 12 $      & \cite{simon2005} \\
$0.120$     & $ \; \qquad 68.6     \pm 26.2$       & \cite{zhang2014} \\
$0.170$     & $ \; \qquad 83     \pm 8$       & \cite{simon2005} \\
$0.179$     & $ \; \qquad 75     \pm 4$       & \cite{moresco2012} \\
$0.199$     & $ \; \qquad 75     \pm 5$        & \cite{moresco2012} \\
$0.200$     & $ \; \qquad 72.9     \pm 29.6$        &  \cite{zhang2014} \\
$0.270$     & $ \; \qquad 77     \pm 14$      & \cite{simon2005} \\
$0.280$     & $ \; \qquad 88.8     \pm 36.6$      & \cite{zhang2014} \\
$0.320$     & $ \; \qquad 79.2   \pm 5.6$     & \cite{cuesta2016}\\
$0.352$     & $ \; \qquad 83     \pm 14$      & \cite{moresco2012} \\
$0.3802$    & $ \; \qquad 83     \pm 13.5$      & \cite{moresco2012} \\
$0.400$     & $ \; \qquad 95     \pm 17$      & \cite{simon2005} \\
$0.4004$    & $ \; \qquad 77     \pm 10.2$      & \cite{moresco2012} \\
$0.4247$    & $ \; \qquad 87.1     \pm 11.2$      & \cite{moresco2012} \\
$0.440$     & $ \; \qquad 82.6   \pm 7.8$     & \cite{blake2012} \\
$0.4497$    & $ \; \qquad 92.8   \pm 12.9$     & \cite{moresco2012} \\
$0.470$     & $ \; \qquad 89   \pm 50$     & \cite{ratsim} \\
$0.4783$    & $ \; \qquad 80.9   \pm 9$     & \cite{moresco2012} \\
$0.480$     & $ \; \qquad 97     \pm 62$      & \cite{stern2010} \\
$0.570$     & $ \; \qquad 100.3  \pm 3.7$     & \cite{cuesta2016} \\
$0.593$     & $ \; \qquad  104   \pm 13$      & \cite{moresco2012} \\
$0.600$     & $ \; \qquad 87.9   \pm 6.1$     & \cite{blake2012} \\
$0.680$     & $ \; \qquad 92     \pm 8$       & \cite{moresco2012} \\
$0.730$     & $ \; \qquad 97.3   \pm 7 $      & \cite{blake2012} \\
$0.781$     & $ \; \qquad 105    \pm 12$      & \cite{moresco2012} \\
$0.875$     & $ \; \qquad 125    \pm 17$      & \cite{moresco2012} \\
$0.880$     & $ \; \qquad 90     \pm 40$      & \cite{stern2010} \\
$0.900$     & $ \; \qquad 117    \pm 23$      & \cite{simon2005} \\
$1.037$     & $ \; \qquad 154    \pm 20 $     & \cite{moresco2012} \\
$1.300$     & $ \; \qquad 168    \pm 17 $     & \cite{simon2005} \\
$1.363$     & $ \; \qquad 160    \pm 33.6$    & \cite{moresco2015}\\
$1.430$     & $ \; \qquad 177    \pm 18$      & \cite{simon2005}\\
$1.530$     & $ \; \qquad 140    \pm 14$      & \cite{simon2005}\\
$1.750$     & $ \; \qquad 202    \pm 40$      & \cite{simon2005}\\
$1.965$     & $ \; \qquad 186.5  \pm 50.4$    & \cite{moresco2015}\\
$2.340$     & $ \; \qquad 222    \pm 7 $      & \cite{delubac2014}\\
$2.360$     & $ \; \qquad 226    \pm  8$      & \cite{font-ribera2014}\\
\hline
\end{tabular}\\
 \end{table}

\end{appendix}

\end{document}